\documentclass[journal,twocolumn,10pt]{IEEEtran} 
\usepackage{epsfig,latexsym}
\usepackage{float}
\usepackage{indentfirst}
\usepackage{graphicx}
\usepackage{float}
\usepackage{amsmath}
\usepackage{amssymb}
\usepackage{times}
\usepackage{subfigure}
\usepackage{algorithm}
\usepackage{algorithmic}
\usepackage{pifont}
\usepackage{psfrag}
\usepackage{cite}
\usepackage{url}
\usepackage{lastpage}
\usepackage{bm}
\usepackage{color}
\usepackage{fancyhdr}
\usepackage[center]{caption2}
\usepackage{balance}
\usepackage{fancyhdr,lastpage}
\usepackage{multirow}
\usepackage{mmap} 

\usepackage{bbding}
\usepackage{amsmath}
\usepackage{amssymb}
\usepackage{ntheorem}
\usepackage{graphicx}
\usepackage{xcolor}

\psfull

\begin{document}
\title{Beamforming and Power Splitting Designs for AN-aided Secure Multi-user MIMO SWIPT Systems}
\author{Zhengyu Zhu, \emph{Student Member, IEEE}, Zheng Chu, Ning Wang, \emph{Member, IEEE},
\\Sai Huang, Zhongyong Wang, and Inkyu Lee, \emph{Fellow, IEEE}

\thanks{Part of this work was presented at the 2016 IEEE International Conference on Communications (ICC'16), May 23-27, 2016, Kuala Lumpur, Malaysia.}
\thanks{Z. Zhu is with the School of Information Engineering, Zhengzhou University, Zhengzhou, China.
He was with the School of Electrical Engineering, Korea University, Seoul, Korea (e-mail: zhuzhengyu6@gmail.com).
Z. Chu  is with the School of Science and Technology, Middlesex University, The Burroughs, London, NW4 4BT, UK. (e-mail: z.chu@ncl.ac.uk).
N. Wang and Z. Wang are with the School of Information Engineering, Zhengzhou University, Zhengzhou 450001, China (e-mail: \{ienwang, iezywang\}@zzu.edu.cn).
S. Huang is with the School of Information and Communication Engineering, Beijing University of Posts and Telecommunications, Beijing, China (e-mail: huangsai@bupt.edu.cn).
I. Lee is with School of Electrical Engineering, Korea University, Seoul, Korea (e-mail: inkyu@korea.ac.kr). 
}
}

\maketitle
\begin{abstract}
In this paper, an energy harvesting scheme for a multi-user multiple-input-multiple-output (MIMO) secrecy channel with artificial noise (AN) transmission is investigated.
Joint optimization of the transmit beamforming matrix, the AN covariance matrix, and the power splitting ratio is conducted to minimize the transmit power under the target secrecy rate, the total transmit power, and the harvested energy constraints.
The original problem is shown to be non-convex, which is tackled by a two-layer decomposition approach.
The inner layer problem is solved through semi-definite relaxation, and the outer problem, on the other hand, is shown to be a single-variable optimization that can be solved by one-dimensional (1-D) line search.
To reduce computational complexity, a sequential parametric convex approximation (SPCA) method is proposed to find a near-optimal solution.
The work is then extended to the imperfect channel state information case with norm-bounded channel errors. 
Furthermore, tightness of the relaxation for the proposed schemes are validated by showing that the optimal solution of the relaxed problem is rank-one.
Simulation results demonstrate that the proposed SPCA method achieves the same performance as the scheme based on 1-D but with much lower complexity.
\end{abstract}
\section{Introduction}
In recent years, the idea of energy harvesting (EH) has been introduced to power electronic devices by energy captured from
the environment. 
However, harvesting from natural energy sources such as solar and wind depends on many factors and thus introduces severe reliability issue.
Radio frequency (RF) signal can be utilized as an alternative to more reliably deliver energy to EH devices while simultaneously transmitting information \cite{1EH,2Ruizhang_WPC,3Ding_Commun_mag,4NG_EH_OFDMA}.
Based on this idea, simultaneous wireless information and power transfer (SWIPT) schemes have been proposed to extend the lifetime
of wireless networks \cite{5ZhangR_13TWC_MIMO_SWIPT,6Zhu_15VTC,7HoonLee_15TWC_SWIPT,8zhu_16VTC,9zhu16_JCN}.
For SWIPT operation in multiple antenna systems \cite{7HoonLee_15TWC_SWIPT,8zhu_16VTC}, co-located receiver architecture employing a power splitter for EH and information decoding (ID) has been studied \cite{9zhu16_JCN}.

On the other hand, in the literature we see increasing research interest in secrecy transmission through physical layer (PHY) security designs \cite{10Poor_10TSP_security}.
Unlike conventional cryptographic methods which are normally adopted in the network layer and rely on computational security, PHY security approaches are developed from the information-theoretic perspective such that provable secrecy capacity can be achieved \cite{11chutwc,12Khisti_10TIT_Secure_multiple_antennas_I,13chu_TVT,SPCA_SWIPT_TVT}.
PHY security techniques have been proposed to enhance information security of multiple antenna systems by casting more interference to potential eavesdroppers. 
By adding artificial noise (AN) and projecting it onto the null space of information user channels in transmit beamforming, the potential eavesdroppers would experience a higher noise floor and thus obtain less information about the messages transmitted to the legitimate receivers \cite{14Negi_08TWC_tichu_AN,15LiQ_13TSP_Spatially_selective_AN}.
In SWIPT systems, AN injection can improve secrecy capacity of information transmission while not affecting simultaneous power transfer \cite{16LiuL_14TSP_Secrecy_WIPT_MISO,17Khan_15TIFS_masked_BF_EH,18Shi_Secrecy_MISO_BC_SWIPT, 19Ng_robust_secureSWIPT,20Robust_AN_Aided_Secure_SWIPT,21Chu_SWIPT_MISO_secrecy,22zhu_16ICC,23XiongJ_15masked_beamforming,SPCA_SWIPT_SPL}. 
The AN-aided beamforming for SWIPT operation has been investigated in various multiple-input multiple-output (MIMO) channels \cite{18Shi_Secrecy_MISO_BC_SWIPT,19Ng_robust_secureSWIPT,20Robust_AN_Aided_Secure_SWIPT,21Chu_SWIPT_MISO_secrecy,22zhu_16ICC,SPCA_SWIPT_SPL}.
More recently, robust AN-aided transmit beamforming with unknown eavesdroppers was studied for multiple-input single-output (MISO)
cognitive radio systems based on different channel uncertainty models \cite{23XiongJ_15masked_beamforming}.

In SWIPT systems, when information receivers (IRs) and energy-harvesting receivers (ERs) are in the same cell, the ERs are normally closer to the  transmitter, compared with the IRs, because the power sensitivity level of ER is typically low.
This raises a new information security issue for SWIPT systems because the ERs can potentially eavesdrop the information transmission to the IRs with relatively higher received signal strength \cite{19Ng_robust_secureSWIPT,22zhu_16ICC,24Schober_14_ICC}.
In order to guarantee information security for the IRs, it is desirable to implement some mechanism to prevent the ERs from recovering the confidential message from their observations.


Motivated by the aforementioned observations, in this paper, we study secrecy transmission over a multi-user MIMO secrecy channel which consists of one multi-antenna transmitter, multiple legitimate single antenna co-located receivers (CRs) and multiple multi-antenna ERs.
We employ an AN injection scheme to mask the desired information-bearing signals for secrecy consideration without imposing any structural restriction on the AN.
In comparison with existing works which do not consider power splitter at the legitimate receivers  \cite{16LiuL_14TSP_Secrecy_WIPT_MISO,17Khan_15TIFS_masked_BF_EH,20Robust_AN_Aided_Secure_SWIPT,21Chu_SWIPT_MISO_secrecy}, in this paper,
each CR is assumed to adopt a power splitter to collect energy from both the information-bearing signal and the AN.
The design objective is to jointly optimize the transmit beamforming matrix, the AN covariance matrix, and the power splitting (PS) ratio
such that the AN transmit power is maximized\footnote{AN power maximization is equivalent to minimizing the transmit power of the information signal \cite{18Shi_Secrecy_MISO_BC_SWIPT}.} subject to constraints on the secrecy rate, the total transmit power, and the energy harvested by both the CRs and the ERs.
Because of the coupling effect in the joint optimization problem, determination of the AN covariance matrix and the PS ratio makes the derivation of the secrecy rate and the harvested energy at the CRs more complicated. 

The formulated power minimization (PM) problem for AN-aided secrecy transmission is shown to be non-convex, which cannot be solved directly \cite{26Boyd_Convex}.
The PM problem is thus transformed into a two-layer optimization problem and solved accordingly through semi-definite relaxation (SDR) and one-dimensional (1D) line search. We first propose a joint optimization design for the case with perfect channel state information (CSI). The framework is then extended to robust designs for systems having deterministic or statistical CSI uncertainties.
The contributions of this work are summarized as follows:
\begin{itemize}
\item For the case with perfect CSI at the transmitter, 
the inner loop of the PM problem is solved through SDR, 
while the outer loop is shown to be a single-variable optimization problem, where a one-dimensional line search algorithm 
is employed to find the optima. To reduce computational complexity, a sequential parametric convex approximation (SPCA) method is also investigated \cite{SPCA_SWIPT_TVT,SPCA_SWIPT_SPL,SPCA_method}. 
\item  For the imperfect CSI case with deterministic channel uncertainties, we consider a worst-case robust PM (WCR-PM) problem.
By exploiting the \emph{S}-procedure \cite{26Boyd_Convex}, the semi-infinite constraints are transformed into linear matrix inequalities (LMIs) and the inner loop can be relaxed into an SDP by employing the SDR method.
The corresponding robust optimal design is proposed. Furthermore, an SPCA-based iterative algorithm is also addressed with low complexity.
\item For both the perfect CSI and imperfect CSI cases, the tightness of the SDP relaxation is verified by showing that the optimal solution is rank-one.

\end{itemize}

Compared with our preliminary work \cite{22zhu_16ICC}, major additional work and results incorporated in this paper are summarized in the following. 1) This paper has extended the problem of AN power maximization to both
perfect and imperfect CSI cases, which introduces substantial changes in the analyses. 2) An SPCA-based iterative algorithm has been proposed to solve the problem such that the computational complexity is largely reduced compared with the 1-D search method used in the previous work.


The rest of this paper is organized as follows:
The system model of a multi-user MIMO secrecy channel with SWIPT is presented in Section II.
Section III investigates the transmit beamforming based PM problem with perfect CSI.
Section IV extends the PM results to the imperfect CSI case. 
Section V illustrates the computational complexity of the proposed algorithms. The numerical results are shown in Section VI. Finally, we conclude the paper in Section VII.

\textbf{\emph{Notation:}} Vectors and matrices are denoted by bold lowercase and bold uppercase letters, respectively.
 $(\cdot)^{T}$ and $(\cdot)^H$ represent matrix transpose and Hermitian transpose. 
 The operator $\otimes$ represents the Kronecker product. For a vector $\emph{\textbf{x}}$, $\|\textbf{\emph{x}}\|$ indicates the
  Euclidean norm. $\mathbb{C}^{M\times L}$ and $\mathbb{H}^{M\times L}$ denote the space of ${M\times L}$ complex matrices
  and Hermitian matrices, respectively. $ \mathbb{H}_{+}$ represents the set of positive semi-definite Hermitian matrices, and
   $ \mathbb{R}_{+}$ denotes the set of all nonnegative real numbers. For a matrix ${\textbf{A}}$, ${\textbf{A}}\succeq \textbf{0}$ means
    that ${\textbf{A}}$ is positive semi-definite, and $\|\textbf{{A}}\|_F$, $\rm{tr} ({\textbf{A}})$, $|{\textbf{A}}|$ and $\rm{rank}({\textbf{A}})$
    denote the Frobenius norm, trace, determinant,  and the rank, respectively. ${{\emph{vec}}}(\textbf{{A}})$ stacks the elements of $\textbf{{A}}$
    in a column vector. $\mathbf{0}_{M\times L}$ is a zero matrix of size ${M\times L}$. $\mathrm{E}\{\cdot\}$ is the expectation operator, and $\Re\{\cdot\}$ stands for the real part of a complex number.
     $[x]^{+}$ represents $\max\{x,0\}$ and $\lambda_{max}(\textbf{{A}})$ denotes the maximum eigenvalue of $\textbf{{A}}$.

\section{System Model}
In this section, we consider a multi-user MIMO secrecy channel which consists of one multi-antenna transmitter,
$ L $ single-antenna CRs and $ K $ multi-antenna ERs. We assume that each CR employs the PS scheme to receive the information and harvest power simultaneously.
It is assumed that the transmitter is equipped with $ N_{T} $ transmit antennas, and each ER has $ N_{R} $ receive antennas.

We denote by $ \mathbf{h}_{c,l} \in \mathbb{C}^{N_{T}} $ the channel vector between the transmitter and the $ l $-th CR, and $ \mathbf{H}_{e,k} \in  \mathbb{C}^{N_{T} \times N_{R}} $ the channel matrix between the transmitter and the $ k $-th ER.
The received signal at the $ l $-th CR and the $ k $-th ER are given by
\begin{equation*}
\begin{split}
y_{c,l} & = \mathbf{h}_{c,l}^{H}\mathbf{x} \!+\! n_{c,l}, ~\forall l,\\
\mathbf{y}_{e,k} & = \mathbf{H}_{e,k}^{H}\mathbf{x}\!+\! \mathbf{n}_{e,k}, ~\forall k,
\end{split}
\end{equation*}
where $ \mathbf{x} \in \mathbb{C}^{N_{T}} $ is the transmitted signal vector, and $ n_{c,l} \sim \mathcal{CN}(0, \sigma_{c,l}^{2}) $
and $ \mathbf{n}_{e,k} \sim \mathcal{CN}(0, \sigma_{k}^{2}\mathbf{I}) $ are the additive Gaussian noise at the $ l $-th CR and the $ k $-th ER, respectively.

In order to achieve secure transmission, the transmitter employs
transmit beamforming with AN, which acts as interference to the ERs, and provides energy to the CRs and ERs.
The transmit signal vector $ \mathbf{x} $ can be written as
\begin{equation}
\mathbf{x} = \mathbf{q}s + \mathbf{w},
\end{equation}
where $ \mathbf{q} \in \mathbb{C}^{N_{T}}$ defines the transmit beamforming vector, $s$ with $\mathrm{E}\{s^2\} = 1$ is the information-bearing signal intended for the CRs, and $ \mathbf{w} \in \mathbb{C}^{N_{T}}$ represents the energy-carrying AN, which can also be composed by multiple energy beams.

As the CR adopts PS to perform ID and EH simultaneously,
the received signal at the $l$-th CR is divided into ID and EH components by the PS ratio $\rho_{c,l}\in (0, 1]$.
Therefore, the signal for information detection at the $l$-th CR is given by
\begin{equation*}
y_{c,l}^{ID}   = \sqrt{\rho_{c,l}}y_{c,l} \!+\! n_{p,l}
 =  \sqrt{\rho_{c,l}}(\mathbf{h}_{c,l}^{H}\mathbf{x} \!+\! n_{c,l}) \!+\! n_{p,l}, ~\forall l,
\end{equation*}
where $ n_{p,l} \sim \mathcal{CN}(0, \sigma_{p,l}^{2}) $  is the additive Gaussian noise at the  $l$-th CR.

Denoting $ \mathbf{Q} = \mathrm{E}\{\mathbf{q}\mathbf{q}^H\} $ as the transmit covariance matrix and
$ \mathbf{W}= \mathrm{E}\{\mathbf{w}\mathbf{w}^H\} $ as the AN covariance matrix,
the achieved secrecy rate at the $ l $-th CR is given by
\begin{equation}\label{eq:Achieved_sec_rate_with_user_PS}
\begin{split}
\hat{R}_{c,l} &= \bigg[\log\bigg( 1 \!+\! \frac{\rho_{c,l} \mathbf{h}_{c,l}^{H}\mathbf{Q}\mathbf{h}_{c,l}}{\rho_{c,l}
(\sigma_{c,l}^{2}\!+\!\mathbf{h}_{c,l}^{H}\mathbf{W}\mathbf{h}_{c,l}) \!+\! \sigma_{p,l}^{2}} \bigg)\\
&- \max_{k} \log \bigg| \mathbf{I} \!+\! (\mathbf{H}_{e,k}^{H}\mathbf{W}\mathbf{H}_{e,k} \!+\!
 \sigma_{k}^{2}\mathbf{I})^{-1} \mathbf{H}_{e,k}^{H}\mathbf{Q}\mathbf{H}_{e,k} \bigg| \bigg]^{+},~\forall l.
\end{split}
\end{equation}
The harvested power at the $ l $-th CR and the $ k $-th ER is therefore
\begin{equation}
\begin{split}
E_{c,l} & = \eta_{c,l}(1 \!-\! \rho_{c,l})\big(\mathbf{h}_{c,l}^{H}(\mathbf{Q}\!+\!\mathbf{W})\mathbf{h}_{c,l} \!+\! \sigma_{c,l}^{2}\big), ~\forall l,\\
E_{e,k} & = \eta_{e,k}\bigg(\textrm{tr}\big(\mathbf{H}_{e,k}^{H}(\mathbf{Q}\!+\!\mathbf{W})\mathbf{H}_{e,k}\big) \!+\! N_{R}\sigma_{k}^{2}\bigg), ~\forall k,
\end{split}
\end{equation}
where $ \eta_{c,l} $ and $ \eta_{e,k} $ represent the EH efficiency of the $ l $-th CR and
the EH efficiency of the $ k $-th ER, respectively. In this paper we set $ \eta_{c,l}=\eta_{e,k}=0.3 $ for simplicity.
The results can be easily extended to scenarios with different $ \eta_{c,l} $ and $ \eta_{e,k} $ values.
In the following section, we consider the transmit beamforming based PM problem
to jointly optimize the transmit covariance matrix $ \mathbf{Q}$, the AN covariance matrix $ \mathbf{W} $, and the PS ratio $ \rho_{c,l} $.

\section{Masked Beamforming Based Power Minimization with Perfect CSI}
In this section, we study transmit beamforming optimization under the assumption that perfect CSI of all the channels is available at the transmitter.
\subsection{Problem Formulation}
In this problem, the transmit power of the information signal is minimized subject to the
 total transmit power constraint, the secrecy rate constraint, and the EH constraints of the CRs and the ERs
 such that the AN transmit power is maximized for secrecy consideration.
The AN-aided PM problem is thus formulated as
\begin{subequations}\label{eq:Masked_beamforming_sec_rate_opt_ori}
\begin{eqnarray}
\min_{\mathbf{Q},{\kern 1pt}\mathbf{W},{\kern 1pt}\rho_{c,l}} &&~~~~~~~~~~~~ \textrm{tr}(\mathbf{Q}) \nonumber\\
\mbox{s.t.} ~~~ &&\!\!\!\!\!\!\!\!\!\!\!\!\!\! \log\bigg( 1 \!+\! \frac{\rho_{c,l} \mathbf{h}_{c,l}^{H}\mathbf{Q}\mathbf{h}_{c,l}}{\rho_{c,l} (\sigma_{c,l}^{2}\!+\!\mathbf{h}_{c,l}^{H}\mathbf{W}\mathbf{h}_{c,l}) \!+\!
\sigma_{p,l}^{2}} \bigg) \nonumber  \\
&&\!\!\!\!\!\!\!\!\!\!\!\!\!\!\!\!\!\!\!\!\!\!\!\!\!\!\!\!\!\!\!\!\!\!\!\!\!\!\!\!\!\!\! - \max_{k} \log \bigg| \mathbf{I} +  (\sigma_{k}^{2}\mathbf{I} \!+\!
\mathbf{H}_{e,k}^{H}\mathbf{W}\mathbf{H}_{e,k})^{-1}\mathbf{H}_{e,k}^{H}\mathbf{Q}\mathbf{H}_{e,k} \bigg| \!\geq\! \bar{R}_{c,l},
\label{eq:Achieved_sec_rate_constraint}\\
&&\!\!\!\!\!\!\!\!\!\!\!\!\!\!\!\!\!\!\!\!\!\!\! \textrm{tr}(\mathbf{Q} + \mathbf{W}) \leq P,  \label{eq:Power_constraints}\\
&&\!\!\!\!\!\!\!\!\!\!\!\!\!\!\!\!\!\!\!\!\!\!\! \mathbf{h}_{c,l}^{H}(\mathbf{Q} + \mathbf{W})\mathbf{h}_{c,l} + \sigma_{c,l}^{2} \geq \frac{\bar{E}_{c,l}}{\eta_{c,l}(1-\rho_{c,l})}, \label{eq:Energy_constraint_user_PS}\\
&&\!\!\!\!\!\!\!\!\!\!\!\!\!\!\!\!\!\!\!\!\!\!\!  \min_{k} \textrm{tr}\big(\mathbf{H}_{e,k}^{H}(\mathbf{Q}\!+\!\mathbf{W})\mathbf{H}_{e,k}\big) \!+\! N_{R}\sigma_{k}^{2}  \geq \frac{\bar{E}_{e,k}}{\eta_{e,k}},\forall k, \label{eq:Energy_constraint_eve}\\
&&\!\!\!\!\!\!\!\!\!\!\!\!\!\!\!\!\!\!\!\!\!\!\! \mathbf{Q}\succeq \mathbf{0},\mathbf{W} \succeq \mathbf{0}, 0 < \rho_{c,l} \leq 1, \forall l,  \textrm{rank}(\mathbf{Q}) \!=\! 1, \label{eq:Another_constraints_ori}
\end{eqnarray}
\end{subequations}
where $ \bar{R}_{c,l} $ is the target secrecy rate, $ P $ is the total transmit power,
and $ \bar{E}_{c,l} $ and $ \bar{E}_{e,k} $ denote the predefined harvested power at the $l$-th CR and
the $k$-th ER, respectively.
The constraint (\ref{eq:Energy_constraint_eve}) guarantees that a minimum energy harvested power should be achieved by the $k$-th ER.
\subsection{One-Dimensional Line Search Method (1-D Search)}
Problem \eqref{eq:Masked_beamforming_sec_rate_opt_ori} is non-convex due to the
 secrecy rate constraint (\ref{eq:Achieved_sec_rate_constraint}), and thus cannot be solved directly.
 In order to circumvent this issue, we convert the original problem
by introducing a slack variable $ t $ for the $ k $-th ER's rate. Then we have
\begin{subequations}\label{eq:relaxed_power_min_problem1}
\begin{eqnarray}
&&\!\!\!\!\!\!\!\!\!\!\!\!\!\!\! \min_{\mathbf{Q},{\kern 1pt}\mathbf{W},{\kern 1pt}\rho_{c,l},{\kern 1pt}t}  ~~~~~~ \textrm{tr}(\mathbf{Q}) \nonumber \\
&&\!\!\!\!\!\!\!\!\!\!\!\!\!\!\! \mbox{s.t.}~~ \!  \log\bigg(t \!+\! \frac{t\rho_{c,l}\mathbf{h}_{c,l}^{H}\mathbf{Q}\mathbf{h}_{c,l}}{\rho_{c,l}(\sigma_{c,l}^{2}
\!+\!\mathbf{h}_{c,l}^{H}\mathbf{W}\mathbf{h}_{c,l})\!+\!\sigma_{p,l}^{2}}\bigg) \!\geq\! \bar{R}_{c,l}, \forall l,\label{eq:objective_function_with_log}
\\
&&\!\!\!\!\!\!\!  \bigg| \mathbf{I} \!+\! (\sigma_{k}^{2}\mathbf{I} \!+\! \mathbf{H}_{e,k}^{H}\mathbf{W}\mathbf{H}_{e,k})^{-1}\mathbf{H}_{e,k}^{H}\mathbf{Q}\mathbf{H}_{e,k} \bigg|
 \! \leq \! \frac{1}{t},\forall k,  \label{eq:Eavesdroppers_rate_slack_variable_log}  \\
&&\!\!\!\!\!\!\! \eqref{eq:Power_constraints}-\eqref{eq:Another_constraints_ori}. \nonumber
\end{eqnarray}
\end{subequations}

Problem \eqref{eq:relaxed_power_min_problem1} is still non-convex in constraints
\eqref{eq:objective_function_with_log} and \eqref{eq:Eavesdroppers_rate_slack_variable_log},
which can be addressed by reformulating \eqref{eq:relaxed_power_min_problem1} into a two-layer problem.
For the inner layer, we solve problem \eqref{eq:relaxed_power_min_problem1} for a given $ t $, which
is relaxed as
\begin{subequations}\label{eq:relaxed_power_min_problem_results}
\begin{eqnarray}
&&\!\!\!\!\!\!\!\!\!\!\!\!\!\!\!\!\!\!\!\!\!\! f(t) = \min_{\mathbf{Q},{\kern 1pt}\mathbf{W},{\kern 1pt}\rho_{c,l},{\kern 1pt}t}  ~~  \textrm{tr}(\mathbf{Q}) \nonumber\\
&&\!\!\!\!\!\!\!\!\!\!\!\!\!\!\!\!\!\!\!\!\!\!  \mbox{s.t.}~    \mathbf{h}_{c,l}^{H}\big( t \mathbf{Q} \!-\! (2^{\bar{R}_{c,l}} \!-\! t)\mathbf{W} \big) \mathbf{h}_{c,l} \!\geq\!
(2^{\bar{R}_{c,l}} \!-\! t)\big(\sigma_{c,l}^{2}\!+\!\frac{\sigma_{p,l}^{2}}{\rho_{c,l}}\big), \label{eq:objective_function_with_log_relaxed}\\
&&\!\!\!\!   ( {\textstyle{1 \over t}} \!-\! 1)(\sigma_{k}^{2}\mathbf{I} \!+\! \mathbf{H}_{e,k}^{H}\mathbf{W}\mathbf{H}_{e,k}) \!\succeq\!
\mathbf{H}_{e,k}^{H}\mathbf{Q}\mathbf{H}_{e,k},\forall k, \label{eq:Eavesdroppers_rate_slack_variable_log_relaxed}\\
&&\!\!\!\!   \eqref{eq:Power_constraints}-\eqref{eq:Another_constraints_ori}, \nonumber
\end{eqnarray}
\end{subequations}
where $ f(t) $ is defined as  the optimal value of problem \eqref{eq:relaxed_power_min_problem_results}, which is a function of $ t $.
Even though the function $f(t)$ cannot be expressed in closed-form, numerical evaluation of $f(t)$ is feasible.

%
%
%
%
%

\emph{Remark 1:} It is noted that the LMI constraint \eqref{eq:Eavesdroppers_rate_slack_variable_log_relaxed} is obtained from \cite[Proposition 1]{15LiQ_13TSP_Spatially_selective_AN}, and is based on the assumption that $ \textrm{rank}(\mathbf{Q}) \leq 1 $, which will be shown later.

By ignoring the non-convex constraint $ \textrm{rank}(\mathbf{Q}) = 1 $, problem \eqref{eq:relaxed_power_min_problem_results} becomes convex and thus can be solved efficiently by an interior-point method for any given $ t $ \cite{26Boyd_Convex}.
The outer layer problem, whose objective is to find the optimal value of $ t $, is then formulated as
\begin{equation}\label{eq:Outer_level_problem_over_t}
\begin{split}
& \min_{t} ~ f(t) \\
& \mbox{s.t.} ~~ t_{\textrm{min}} \!\leq\! t \!\leq\! t_{\textrm{max}},
\end{split}
\end{equation}
where $t_{\textrm{max}}$ and $t_{\textrm{min}}$ are the upper and lower bounds of $t$, respectively.
The solution to problem \eqref{eq:Outer_level_problem_over_t} can be found by one-dimensional line search. 
For the line search algorithm, we need to determine the lower and upper bounds of the searching interval for $ t $.
It is straightforward that $ t_{\textrm{max}} = 1 $ can be used as the upper bound due to the feasibility of \eqref{eq:Eavesdroppers_rate_slack_variable_log}, while a lower bound is calculated as
\begin{eqnarray}
t &&\geq \min_{l}  \bigg(1 + \frac{\rho_{c,l} \mathbf{h}_{c,l}^{H}\mathbf{Q}\mathbf{h}_{c,l}}{\rho_{c,l} (\sigma_{c,l}^{2}+\mathbf{h}_{c,l}^{H}\mathbf{W}\mathbf{h}_{c,l}) + \sigma_{p,l}^{2}} \bigg)^{-1} \nonumber\\
 &&\geq  \min_{l} \bigg(1 +\frac{\mathbf{h}_{c,l}^{H}\mathbf{Q}\mathbf{h}_{c,l}}{\sigma_{c,l}^{2}+\sigma_{p,l}^{2}+\mathbf{h}_{c,l}^{H}\mathbf{W}\mathbf{h}_{c,l}} \bigg)^{-1} \\
 &&\geq  \min_{l} \bigg(1+\frac{P \|\mathbf{h}_{c,l}\|^{2}}{\sigma_{c,l}^{2}+\sigma_{p,l}^{2}}\bigg)^{-1} \triangleq t_{\textrm{min}}\nonumber.
\end{eqnarray}
where the first inequality is based on the secrecy rate $\bar{R}_{c,l} \geq 0$, the third inequality
follows from \eqref{eq:Power_constraints}.
In the following theorem, we prove the equivalence of problem \eqref{eq:Outer_level_problem_over_t} and the original problem
\eqref{eq:Masked_beamforming_sec_rate_opt_ori}.

\emph{\underline{Theorem 1:}}
The transmit beamforming PM problem
\eqref{eq:Masked_beamforming_sec_rate_opt_ori} is equivalent to problem
\eqref{eq:Outer_level_problem_over_t} when $ \textrm{rank}(\mathbf{Q}) \leq 1 $.

\emph{\underline{Proof:}}
Let us denote the optimal solutions of \eqref{eq:Masked_beamforming_sec_rate_opt_ori} and \eqref{eq:Outer_level_problem_over_t} as $ f^{*} $ and $ f^{\rm{opt}} $, respectively.
First, we show that $f^{\rm{opt}}$ is a feasible point of problem \eqref{eq:Outer_level_problem_over_t}, 
i.e. $ f^{\rm{opt}} \leq f^{*} $. It is noted that
\eqref{eq:Masked_beamforming_sec_rate_opt_ori} and \eqref{eq:relaxed_power_min_problem_results} have the same objective function and the optimal
solution of \eqref{eq:Masked_beamforming_sec_rate_opt_ori} satisfies the constraints
of \eqref{eq:relaxed_power_min_problem_results} given the assumption
 that $ \textrm{rank}(\mathbf{Q}) \leq 1 $ \cite{15LiQ_13TSP_Spatially_selective_AN}, which gives rise to
  $ f(t^{*}) = f^{*} $, where $t^{*}$ is the optimal value of $t$. In addition, it follows
  $ f^{\rm{opt}} \leq f(t^{*}) $.
  Next, we prove that the solution to problem
   \eqref{eq:relaxed_power_min_problem_results} is achievable in problem \eqref{eq:Masked_beamforming_sec_rate_opt_ori}, i.e. $ f^{*} \leq f^{\rm{opt}} $.
  From \eqref{eq:Achieved_sec_rate_constraint}, \eqref{eq:objective_function_with_log_relaxed} and  \eqref{eq:Eavesdroppers_rate_slack_variable_log_relaxed},
   we can show that the optimal solutions of \eqref{eq:relaxed_power_min_problem_results} are feasible solutions of
  \eqref{eq:Achieved_sec_rate_constraint} when $ \textrm{rank}(\mathbf{Q}) \leq 1 $. Therefore, we conclude that $ f^{*} = f^{\rm{opt}} $.    ~~~~~~~~~~~~~~~~~~~~~~~~~~~~~~~~~~~~~~~~~~~~~~~~~~~~~~~~~~$\blacksquare $

Utilizing the results in Remark 1 and Theorem 1, next we show the tightness of the AN-aided PM problem \eqref{eq:Masked_beamforming_sec_rate_opt_ori}
by the following theorem.

\emph{\underline{Theorem 2:}} Provided that problem
\eqref{eq:relaxed_power_min_problem_results} is feasible for a given $ t > 0 $, there
exists an optimal solution to \eqref{eq:Masked_beamforming_sec_rate_opt_ori} such that the rank of $ \mathbf{Q} $ is always equal to 1.

\emph{\underline{Proof:} } See Appendix A. ~~~~~~~~~~~~~~~~~~~~~~~~~~~~~~~~~~~~~~~~~~~~~~~~~~~~~~~$\blacksquare $

Problem \eqref{eq:Outer_level_problem_over_t} can be solved by conducting
one-dimensional line search for $f(t)$ over $t$ and choosing the minimum $f(t)$ as the optimal solution. 
Solving the SDP problem \eqref{eq:relaxed_power_min_problem_results} with the optimal $f(t)$, we can obtain the optimal design variables $(\mathbf{Q}^*,{\kern 1pt}\mathbf{W}^*,{\kern 1pt}\rho_{c,l}^*)$. The optimal beamforming vector $\mathbf{q}^*$ is then computed by eigenvalue decomposition $\mathbf{Q}^* = \mathbf{q}^*\mathbf{q}^{*H}$.
\subsection{Low-Complexity SPCA Algorithm}
In this subsection, we propose an SPCA based iterative method to reduce the computational complexity.
By introducing two slack variables $r_1 > 0$ and $r_2 > 0$, the constraint \eqref{eq:Achieved_sec_rate_constraint} can be rewritten as
\begin{subequations}\label{eq:9}
\begin{eqnarray}
&&\!\!\!\!\!\!\!\!\!\!\!\!\!\!  \log( r_1r_2 ) \geq  \bar{R}_{c,l}, \forall l,\label{eq:9a}\\
&&\!\!\!\!\!\!\!\!\!\!\!\!\!\!  1 \!+\! \frac{\rho_{c,l} \mathbf{h}_{c,l}^{H}\mathbf{Q}\mathbf{h}_{c,l}}{\rho_{c,l} (\sigma_{c,l}^{2}\!+\!\mathbf{h}_{c,l}^{H}\mathbf{W}\mathbf{h}_{c,l}) \!+\!
\sigma_{p,l}^{2}}  \geq r_1, \forall l, \label{eq:9b}\\
&&\!\!\!\!\!\!\!\!\!\!\!\!\!\! 1 + \frac{{\rm{tr}}({{{{{{ {\mathbf{H}}}}^H_{e,k}}}}
 {{{\mathbf{Q}}}}{{{ {\mathbf{H}}}}_{e,k}}})}
 {{\sigma _k^2 + {\rm{tr}}({{{{{ {\mathbf{H}}}}^H_{e,k}}}}{\mathbf{W}}{{{ {\mathbf{H}}}}_{e,k}}})}  \leq \frac{1}{r_2}, \forall k, \label{eq:9c}
\end{eqnarray}
\end{subequations}
which can be further simplified as
\begin{subequations} \label{eq:10}
\begin{eqnarray}
&&\!\!\!\!\!\!\!\!\!\!\!\!\!\!   r_1r_2  \geq  2^{\bar{R}_{c,l}}, \forall l, \label{eq:10a}\\
&&\!\!\!\!\!\!\!\!\!\!\!\!\!\!   \frac{ \mathbf{h}_{c,l}^{H}\mathbf{Q}\mathbf{h}_{c,l}}{ \sigma_{c,l}^{2}\!+\!\mathbf{h}_{c,l}^{H}\mathbf{W}\mathbf{h}_{c,l} \!+\!
\frac{\sigma_{p,l}^{2}}{\rho_{c,l}}}  \geq {r_1} -1, \forall l,\label{eq:10b}\\
&&\!\!\!\!\!\!\!\!\!\!\!\!\!\!  \frac{{\sigma _k^2 + {{\rm{tr}}({{{{{{ {\mathbf{H}}}}^H_{e,k}}}}
 {{{\mathbf{W}}}}{{{ {\mathbf{H}}}}_{e,k}}})}}}
 {{{\sigma _k^2}}+{{\rm{tr}}({{{{{{ {\mathbf{H}}}}^H_{e,k}}}}
 ({{{\mathbf{Q}}}}+{{{\mathbf{W}}}}){{{ {\mathbf{H}}}}_{e,k}}})}}  \geq  {{r_2}}, \forall k. \label{eq:10c}
\end{eqnarray}
\end{subequations}
The inequality constraint \eqref{eq:10a} is equivalent to ${2^{\bar{R}_{c,l}+2}} + (r_1-r_2)^2 \leq (r_1+r_2)^2$,
which can be converted into a conic quadratic-representable function form as
\begin{equation} \label{eq:11a}
\left\|\left[\sqrt{2^{\bar{R}_{c,l}+2}}~~~~ r_1-r_2 \right]\right\|\leq r_1+r_2, \forall l.
\end{equation}

By transforming inequality constraints \eqref{eq:10b} and \eqref{eq:10c} into
\begin{subequations}\label{eq:13}
\begin{eqnarray}
&&\!\!\!\!\!\!\!\!   {\sigma_{c,l}^{2}\!+\!\mathbf{w}^{H}\mathbf{H}_{c,l}\mathbf{w} \!+\!
\frac{\sigma_{p,l}^{2}}{\rho_{c,l}}} \leq \frac{\mathbf{q}^H{\mathbf{H}_{c,l}}\mathbf{q}}{r_1 -1}, ~\forall l, \label{eq:13a}\\
&&\!\!\!\!\!\!\!\!\!\!\!\!\!\!\!\!\!\!\!\!\!\!\!\!\!   {{{\sigma _k^2}}+ \mathbf{w}^{H}\mathbf{\hat{H}}_{e,k}\mathbf{w} +\mathbf{q}^{H}\mathbf{\hat{H}}_{e,k}\mathbf{q} } \leq \frac{{{\sigma _k^2 + \mathbf{w}^{H}\mathbf{\hat{H}}_{e,k}\mathbf{w}}}}{r_2}, ~\forall k,\label{eq:13b}
\end{eqnarray}
\end{subequations}
where $\mathbf{H}_{c,l} = {\mathbf{h}_{c,l}\mathbf{h}_{c,l}^{H}}$ and $\mathbf{\hat{H}}_{e,k} = {\mathbf{H}_{e,k}\mathbf{H}_{e,k}^{H}}$,
we observe that these two constraints are non-convex, but the right-hand side (RHS) of both (\ref{eq:13a}) and (\ref{eq:13b}) have the function form of quadratic-over-linear, which are convex functions \cite{26Boyd_Convex}.
Based on the idea of the constrained convex procedure \cite{SPCA_firstorder},
these quadratic-over-linear functions can be replaced by their first-order expansions, which transforms the problem into convex programming.
Specifically, we define
\begin{eqnarray}\label{eq:zzy1}
f_{\mathbf{{A}},a}(\mathbf{w},t) = \frac{{{\mathbf{w}^{H}\mathbf{{A}}\mathbf{w}}}}{t-a},
\end{eqnarray}
where $\mathbf{{A}} \succeq \textbf{0}$ and $t \geq a$. At a certain point $(\mathbf{\tilde{w}}, \tilde{t})$, the first-order Taylor expansion of \eqref{eq:zzy1}
 is given by
\begin{eqnarray}\label{eq:zzy2}
F_{\mathbf{{A}},a}(\mathbf{w},t,\mathbf{\tilde{w}},\tilde{t}) = \frac{2\Re{\{\mathbf{\tilde{w}}^{H}\mathbf{{A}}\mathbf{w}\}}}{\tilde{t}-a} - \frac{{{\mathbf{\tilde{w}}^{H}\mathbf{{A}}\mathbf{\tilde{w}}}}}{(\tilde{t}-a)^2}(t-a).
\end{eqnarray}
By using the above results of Taylor expansion, for the points $(\mathbf{\tilde{q}}, \tilde{r}_1)$ and $(\mathbf{\tilde{w}}, \tilde{r}_2)$, we can transform constraints \eqref{eq:13a} and \eqref{eq:13b} into convex forms,  respectively, as 
\begin{subequations}\label{eq:zzy3}
\begin{eqnarray}
&&\!\!\!\!\!\!\!\!\!\!\!\!\!\!\!\!\!\!\!    {\sigma_{c,l}^{2}+\mathbf{w}^{H}\mathbf{H}_{c,l}\mathbf{w} +
\frac{\sigma_{p,l}^{2}}{\rho_{c,l}}} \leq F_{{\mathbf{H}_{c,l}},1}(\mathbf{q},r_1,\mathbf{\tilde{q}},\tilde{r}_1), \forall l, \label{eq:zzy3a}\\
&&\!\!\!\!\!\!\!\!\!\!\!\!\!\!\!\!\!\!\!   {{{\sigma _k^2}}+ \mathbf{w}^{H}\mathbf{\hat{H}}_{e,k}\mathbf{w} + \mathbf{q}^{H}\mathbf{\hat{H}}_{e,k}\mathbf{q} } \leq  \sigma _k^2(\frac{2}{\tilde{r}_2} - \frac{r_2}{\tilde{r}^2_2}) \nonumber \\
&&~~~~~~~~~~~~~~~~~~~  +  F_{{\mathbf{\hat{H}}_{e,k}},0}(\mathbf{w},r_2,\mathbf{\tilde{w}},\tilde{r}_2), \forall k.\label{eq:zzy3b}
\end{eqnarray}
\end{subequations}
Denoting $g_{r_1,l} = F_{{\mathbf{H}_{c,l}},1}(\mathbf{q},r_1,\mathbf{\tilde{q}},\tilde{r}_1) -\sigma_{c,l}^{2}- \frac{\sigma_{p,l}^{2}}{\rho_{c,l}}$ and $g_{r_2,k} =  \sigma _k^2(\frac{2}{\tilde{r}_2}\! -\! \frac{r_2}{\tilde{r}^2_2})\! + \! F_{{\mathbf{\hat{H}}_{e,k}},0}(\mathbf{w},r_2,\mathbf{\tilde{w}},\tilde{r}_2) - {{\sigma _k^2}}$, 
\eqref{eq:zzy3a} and \eqref{eq:zzy3b} can be recast as the following second-order cone (SOC) constraints
\begin{subequations}\label{eq:zzy8}
\begin{eqnarray}
&&\!\!\!\!\!\!\!\!\!\!\!\!\!\!\!\!   \big\| [2\mathbf{w}^{H}\mathbf{h}_{c,l}, g_{r_1,l} -1  ]^T \big\|\leq g_{r_1,l} +1, ~\forall l, \label{eq:zzy8a}\\
&&\!\!\!\!\!\!\!\!\!\!\!\!\!\!\!\!\!\!\!\!\!\!\!\!\!    \big\| [2\mathbf{w}^{H}\mathbf{{H}}_{e,k}; 2\mathbf{q}^{H}\mathbf{{H}}_{e,k}; g_{r_2,k} -1]^T \big\| \leq  g_{r_2,k} +1, ~\forall k.\label{eq:zzy8b}
\end{eqnarray}
\end{subequations}
~~Next we employ the SPCA technique for the SOC constraints \eqref{eq:Energy_constraint_user_PS} and \eqref{eq:Energy_constraint_eve} \cite{Complexity_zhu} to obtain convex approximations.
By substituting $\mathbf{q} \triangleq {\mathbf{\tilde{q}}}+\Delta \mathbf{q}$ and $\mathbf{w} \triangleq {\mathbf{\tilde{w}}}+\Delta \mathbf{w}$  into the left-hand side (LHS) of \eqref{eq:Energy_constraint_user_PS}, we obtain
\begin{equation}\label{eq: zzy4}
\begin{split}
&~~~~\mathbf{q}^H\mathbf{H}_{c,l}\mathbf{q}+\mathbf{w}^H\mathbf{H}_{c,l}\mathbf{w} + {{\sigma _{c,l}^2}} \\
&=  ({\mathbf{\tilde{q}}}\!+\!\Delta \mathbf{q})^H\mathbf{H}_{c,l}({\mathbf{\tilde{q}}}\!+\!\Delta \mathbf{q})\!+\!({\mathbf{\tilde{w}}}\!+\!\Delta \mathbf{w})^H\mathbf{H}_{c,l}({\mathbf{\tilde{w}}}\!+\!\Delta \mathbf{w})+{{\sigma _{c,l}^2}} \\
& \ge {\mathbf{\tilde{q}}}^H\mathbf{H}_{c,l}{\mathbf{\tilde{q}}} + 2\Re \{ {\mathbf{\tilde{q}}}^H\mathbf{H}_{c,l}\Delta {\mathbf{q}}\}  + {\mathbf{\tilde{w}}}^H\mathbf{H}_{c,l}{\mathbf{\tilde{w}}} \\
&~~~~ + 2\Re \{ {\mathbf{\tilde{w}}}^H\mathbf{H}_{c,l}\Delta {\mathbf{w}}\}+{{\sigma _{c,l}^2}},
\end{split}
\end{equation}
where the inequality is given by dropping the quadratic terms $\Delta \mathbf{q}^H \mathbf{H}_{c,l}\Delta \mathbf{q}$
 and $\Delta \mathbf{w}^H \mathbf{H}_{c,l}\Delta \mathbf{w}$.
Similarly, in the LHS of \eqref{eq:Energy_constraint_eve}, we have
\begin{equation}\label{eq: zzy5}
\begin{split}
&\textrm{tr}\big(\mathbf{H}_{e,k}^{H}(\mathbf{Q}\!+\!\mathbf{W})\mathbf{H}_{e,k}\big)\\
=& \mathbf{q}^H\mathbf{\hat{H}}_{e,k}\mathbf{q}+\mathbf{w}^H\mathbf{\hat{H}}_{e,k}\mathbf{w} \\
=&  ({\mathbf{\tilde{q}}}\!+\!\Delta \mathbf{q})^H\mathbf{\hat{H}}_{e,k}({\mathbf{\tilde{q}}}\!+\!\Delta \mathbf{q})\!+\!({\mathbf{\tilde{w}}}\!+\!\Delta \mathbf{w})^H\mathbf{\hat{H}}_{e,k}({\mathbf{\tilde{w}}}\!+\!\Delta \mathbf{w}) \\
\ge &  {\mathbf{\tilde{q}}}^H\mathbf{\hat{H}}_{e,k}{\mathbf{\tilde{q}}} \!+\! 2\Re \{ {\mathbf{\tilde{q}}}^H\mathbf{\hat{H}}_{e,k}\Delta {\mathbf{q}}\} +{\mathbf{\tilde{w}}}^H\mathbf{\hat{H}}_{e,k}{\mathbf{\tilde{w}}} \\
 &+ 2\Re \{ {\mathbf{\tilde{w}}}^H\mathbf{\hat{H}}_{e,k}\Delta {\mathbf{w}}\}.
\end{split}
\end{equation}
~~According to \eqref{eq: zzy4} and \eqref{eq: zzy5}, we obtain linear approximations of
 the concave constraints \eqref{eq:Energy_constraint_user_PS} and \eqref{eq:Energy_constraint_eve} as
\begin{equation}\label{eq: zzy6}
\begin{split}
&{\mathbf{\tilde{q}}}^H\mathbf{H}_{c,l}{\mathbf{\tilde{q}}} \!+\! 2\Re \{ {\mathbf{\tilde{q}}}^H\mathbf{H}_{c,l}\Delta {\mathbf{q}}\} \!+\!
{\mathbf{\tilde{w}}}^H\mathbf{H}_{c,l}{\mathbf{\tilde{w}}}\\
&+ 2\Re \{ {\mathbf{\tilde{w}}}^H\mathbf{H}_{c,l}\Delta {\mathbf{w}}\} + \sigma_{c,l}^{2} \geq \frac{\bar{E}_{c,l}}{\eta_{c,l}(1-\rho_{c,l})}, ~\forall l,
\end{split}
\end{equation}
and
\begin{equation}\label{eq: zzy7}
\begin{split}
&{\mathbf{\tilde{q}}}^H\mathbf{\hat{H}}_{e,k}{\mathbf{\tilde{q}}} \!+\! 2\Re \{ {\mathbf{\tilde{q}}}^H\mathbf{\hat{H}}_{e,k}\Delta {\mathbf{q}}\} \!+\!{\mathbf{\tilde{w}}}^H\mathbf{\hat{H}}_{e,k}{\mathbf{\tilde{w}}} \\
&+ 2\Re \{ {\mathbf{\tilde{w}}}^H\mathbf{\hat{H}}_{e,k}\Delta {\mathbf{w}}\} \!+\! N_{R}\sigma_{k}^{2}  \geq \frac{\bar{E}_{e,k}}{\eta_{e,k}}, ~\forall k.
\end{split}
\end{equation}
Finally, by rearranging \eqref{eq:Power_constraints} as
 \begin{equation}\label{eq: zzy8}
\|[{\mathbf{q}}^T~~~{\mathbf{w}}^T]\|\leq \sqrt{P},
\end{equation}
the original problem \eqref{eq:Masked_beamforming_sec_rate_opt_ori} is transformed into
\begin{equation}\label{16}
\begin{split}
&\!\!\!\! \min_{\mathbf{q},{\kern 1pt}\mathbf{w},{\kern 1pt}\rho_{c,l}, {\kern 1pt}r_1, {\kern 1pt}r_2, {\kern 1pt}g_{r_1,l}, {\kern 1pt}g_{r_2,k}}  ~~~~~~ \|\mathbf{q}\| \\
&\!\!\!\! \mbox{s.t.} {\kern 3pt}\eqref{eq:11a}, {\kern 1pt}\eqref{eq:zzy8}, {\kern 1pt}\eqref{eq: zzy6}, {\kern 1pt}\eqref{eq: zzy7}, {\kern 1pt}\eqref{eq: zzy8}, 0 < \rho_{c,l} \leq 1, \forall l.
\end{split}
\end{equation}
Given $\mathbf{\tilde{q}}$, $\mathbf{\tilde{w}}$, $\tilde{r}_1$, and $\tilde{r}_2$, problem \eqref{16} is convex and can be efficiently solved by convex optimization software tools such as CVX \cite{28CVX}.
Based on the SPCA method, an approximation with the current optimal solution can be updated iteratively,
which implies that \eqref{eq:Masked_beamforming_sec_rate_opt_ori} is optimally solved.
In Section \ref{six}, we will show that the proposed SPCA method achieves the same performance as the 1-D search scheme, but with much lower complexity.

\section{Masked Beamfomring Based Robust PM for Imperfect CSI}
Due to channel estimation and quantization errors, it may not be possible to have perfect CSI in practice.
In this section, we extend the PM optimization method to more practical scenarios with imperfect CSI.
Specifically, we consider an AN-aided WCR-PM formulation under norm-bounded channel uncertainty.
\subsection{Worst-Case Based Robust PM Problem}
Now, we adopt imperfect CSI based on the deterministic model \cite{25Lee_imperfectCSI}.
Specifically, the actual channel between the transmitter and the $ l $-th CR, denoted by $ \mathbf{h}_{c,l} $,
and the actual channel between the transmitter and the $ k $-th ER, denoted by $\mathbf{H}_{e,k}$, can be modeled as
\begin{equation} \label{p9}
\begin{split}
\mathbf{{h}}_{c,l} & =  \mathbf{\bar{h}}_{c,l} + \mathbf{e}_{c,l}, \forall l, \\
\mathbf{{H}}_{e,k} & = \mathbf{\bar{H}}_{e,k} + \mathbf{E}_{e,k}, \forall k,
\end{split}
\end{equation}
where $ \mathbf{\bar{h}}_{c,l} $ and $ \mathbf{\bar{H}}_{e,k} $ denote the estimated channel available at the transmitter,
and $ \mathbf{e}_{c,l} $ and $ \mathbf{E}_{e,k} $ are the bounded CSI errors with $ \|\mathbf{e}_{c,l}\| \leq \varepsilon_{c,l} $ and
$ \|\mathbf{E}_{e,k}\|_{F} \leq \varepsilon_{e,k} $, respectively.

We define $\hat{E}_{c,l}\triangleq\frac{\bar{E}_{c,l}}{\eta_{c,l}}$ and ${\hat{E}}_{e,k} \triangleq \frac{\bar{E}_{e,k}}{\eta_{e,k}}$. By taking the CSI model \eqref{p9} into account, the AN-aided WCR-PM problem can be rewritten  as
\begin{subequations}\label{eq:Masked_beamforming_robust_sec_rate_max_problem_ori}
\begin{eqnarray}
 \min_{\mathbf{Q},{\kern 1pt}\mathbf{W},{\kern 1pt}\rho_{c,l}} &&  ~\textrm{tr}(\mathbf{Q}) ~~~  \nonumber\\
&&\!\!\!\!\!\!\!\!\!\!\!\!\!\!\!\!\!\!\!\!\!\!\!\!\!\! \mbox{s.t.} ~ \log\bigg(1 \!+\!\frac{\rho_{c,l}(\mathbf{\bar{h}}_{c,l}\!+\!\mathbf{e}_{c,l})^{H}\mathbf{Q}(\mathbf{\bar{h}}_{c,l}\!+\!\mathbf{e}_{c,l})}{\rho_{c,l}(\sigma_{c,l}^{2}\!+\!
(\mathbf{\bar{h}}_{c,l}\!+\!\mathbf{e}_{c,l})^{H}\mathbf{W}(\mathbf{\bar{h}}_{c,l}\!+\!\mathbf{e}_{c,l}))\!+\!\sigma_{p,l}^{2}}\bigg) \nonumber\\
&& \!\!\!\!\!\!\!\!\!\!\!\!\!\!\!\!\!\!\!\!\!\!\!\!\!\!\!\!\!\!\!\!\!\! - \mathop {\max }\limits_{k} \log\bigg| \mathbf{I} \!+\! {\mathbf{\bar{H}}_E}(\mathbf{\bar{H}}_{e,k}\!+\!\mathbf{E}_{e,k})^{H}\mathbf{Q}(\mathbf{\bar{H}}_{e,k}\!+\!
\mathbf{E}_{e,k}) \bigg| \!\geq\! \bar{R}_{c,l}, \label{eq:Robust_sec_rate_constraint}\\
 &&\!\!\!\!\!\!\!\!\!\!\!\!\!\!\!\!\!\!\!\!\!\!  \textrm{tr}(\mathbf{Q} \!+\! \mathbf{W}) \!\leq\! P, \label{eq:Power_constraints_with_channel_uncertainties}\\
&&\!\!\!\!\!\!\!\!\!\!\!\!\!\!\!\!\!\!\!\!\!\!\!  (\mathbf{\bar{h}}_{c,l}\!+\!\mathbf{e}_{c,l})^{H}(\mathbf{Q} \!+\! \mathbf{W})
(\mathbf{\bar{h}}_{c,l}\!+\!\mathbf{e}_{c,l}) \!+\!
\sigma_{c,l}^{2} \!\geq\! \frac{\hat{E}_{c,l}}{1\!-\!\rho_{c,l}}, \label{eq:Robust_harvested_energy_constraint_user}\\
&&\!\!\!\!\!\!\!\!\!\!\!\!\!\!\!\!\!\!\!\!\!\!\!   \textrm{tr}\big( (\mathbf{\bar{H}}_{e,k}\!+\!\mathbf{E}_{e,k})^{H} (\mathbf{Q}\!+\!\mathbf{W}) (\mathbf{\bar{H}}_{e,k}\!+\!\mathbf{E}_{e,k}) \big)\nonumber\\
&&\!\!\!\!\!\!\!\!\!\!\!\!\!\!\!\!\!\!\!\!\!\!\! ~ \geq {\hat{E}_{e,k}} \!-\! N_{R}\sigma_{k}^{2}, \label{eq:Robust_harvested_energy_constraint_eve}\\
&&\!\!\!\!\!\!\!\!\!\!\!\!\!\!\!\!\!\!\!\!\!\!\! \mathbf{Q} \!\succeq\! \mathbf{0},~\mathbf{W} \!\succeq\! \mathbf{0},~ 0 < \rho_{c,l} \leq 1,~ \textrm{rank}(\mathbf{Q}) \!=\! 1,
\end{eqnarray}
\end{subequations}
where $ \mathbf{\bar{H}}_E = \bigg(\sigma_{k}^{2}\mathbf{I}\!+\!(\mathbf{\bar{H}}_{e,k}\!+\!\mathbf{E}_{e,k})^{H}\mathbf{W}
(\mathbf{\bar{H}}_{e,k}\!+\!\mathbf{E}_{e,k})\bigg)^{-1}$.

\subsection{One-Dimensional Line Search Method}
The above robust PM problem is not convex in terms of the channel uncertainties.
We therefore consider relaxation for constraint \eqref{eq:Robust_sec_rate_constraint} by
introducing a slack variable $ t_1 $ similar to the previous section.
The constraint \eqref{eq:Robust_sec_rate_constraint} is then transformed into
\begin{subequations}\label{eq:Robust_sec_rate_constraint_modified}
\begin{eqnarray}
&&\!\!\!\!\!\!\!\!\!\!\!\!\!\!\!\!\!\!\!\!\!\!(\mathbf{\bar{h}}_{c,l}\!+\!\mathbf{e}_{c,l})^{H}\tilde{\mathbf{T}}_l
 (\mathbf{\bar{h}}_{c,l}\!+\!\mathbf{e}_{c,l}) \geq (2^{\bar{R}_{c,l}} \!-\! t_1)\big(\sigma_{c,l}^{2}\!+\!\frac{\sigma_{p,l}^{2}}{\rho_{c,l}}\big), \label{11a}\\
&&\!\!\!\!\!\!\!\!\!\!\!\!\!\!\!\!\!\!\!\!\!\! ({\textstyle{1 \over t_1}} \!-\! 1)\big(\sigma_{k}^{2}\mathbf{I}\!+\!(\mathbf{H}_{e,k}\!+\!\mathbf{E}_{e,k})^{H}\mathbf{W}(\mathbf{H}_{e,k}\!+\!\mathbf{E}_{e,k})\big) \nonumber \\
&&\!\!\!\!\!\!\!\!\!\!\!\!\!\!\!\!\!\!\!\! \succeq  \big(\mathbf{H}_{e,k}\!+\!\mathbf{E}_{e,k}\big)^{H}\mathbf{Q}\big(\mathbf{H}_{e,k}\!+\!\mathbf{E}_{e,k}\big),~ \forall k, \label{11b}
\end{eqnarray}
\end{subequations}
where $\tilde{\mathbf{T}}_l = t_1\mathbf{Q} \!-\! (2^{\bar{R}_{c,l}} \!-\! t_1)\mathbf{W}$.
The constraint  \eqref{11b} is obtained under the assumption that
$ \textrm{rank}(\mathbf{Q}) \leq 1 $ \cite[Proposition 1]{15LiQ_13TSP_Spatially_selective_AN}.
Problem \eqref{eq:Masked_beamforming_robust_sec_rate_max_problem_ori} has semi-infinite constraint
 \eqref{eq:Robust_harvested_energy_constraint_user}, \eqref{eq:Robust_harvested_energy_constraint_eve}, \eqref{11a}, and \eqref{11b}.
In order to make the problem tractable, we exploit the
\emph{S}-procedure \cite{26Boyd_Convex} to transform the constraints \eqref{eq:Robust_harvested_energy_constraint_user}, \eqref{eq:Robust_harvested_energy_constraint_eve}, \eqref{11a}, and \eqref{11b} into LMIs.
For completeness, the \emph{S}-procedure is presented in Lemma 1 in the following.

{\underline{\emph{Lemma 1:}}} ($\mathcal{S}$-Procedure \cite[Appendix B.2]{26Boyd_Convex}) Let a
function ${\textbf{\emph{f}}_m}({\rm{\emph{\textbf{x}}}})$ with  ${\rm{\textbf{\emph{x}}}} \in \mathbb{C}{^{N \times 1}} (m = 1,2)$ be defined as
\begin{equation}
{\textbf{\emph{f}}_m}({\rm{\emph{\textbf{x}}}})={{\rm{\emph{\textbf{x}}}}^H}{{\rm{\textbf{A}}}_m}{\rm{\emph{\textbf{x}}}} +
2{\mathop{\rm Re}\nolimits} \left\{ {{\rm{\emph{\textbf{b}}}}_m^H{\rm{\textbf{\emph{x}}}}}
\right\} + {\emph{\textbf{c}}_m}
\end{equation}
where ${{\rm{\textbf{A}}}_m} \in {\mathbb{H}^{N\times N}}$, ${{\rm{\emph{\textbf{b}}}}_m} \in {\mathbb{C}^{N \times 1}}$ and ${\emph{\textbf{c}}_m} \in {\mathbb{R}^{N \times 1}}$. Then, ${\textbf{\emph{f}}_m}({\rm{\emph{\textbf{x}}}}) \le 0 $ holds if and only if there exists $\theta  \ge 0$ such that
\begin{equation*}
\theta \left[ {\begin{array}{*{20}{c}}
{{{\rm{\textbf{A}}}_1}}&{{{\rm{\emph{\textbf{b}}}}_1}}\\
{{\rm{\emph{\textbf{b}}}}_1^H}&{{\emph{\textbf{c}}_1}}
\end{array}} \right] - \left[ {\begin{array}{*{20}{c}}
{{{\rm{\textbf{A}}}_2}}&{{{\rm{\textbf{\emph{b}}}}_2}}\\
{{\rm{\emph{\textbf{b}}}}_2^H}&{{\emph{\textbf{c}}_2}}
\end{array}} \right]\succeq {\textbf{0}},
\end{equation*}
provided that there is a point ${\rm{\hat {\emph{\textbf{x}}}}}$ which satisfies
${\emph{\textbf{f}}_m}({\rm{\hat {\emph{\textbf{x}}}}}) < 0$. $\blacksquare$

To employ the \emph{S}-procedure, we rewrite the first constraint in \eqref{eq:Robust_sec_rate_constraint_modified} as
\begin{equation}\label{eq:Robust_sec_rate_constraint_modified zhu}
\mathbf{e}_{c,l}^{H}\tilde{\mathbf{T}}_l \mathbf{e}_{c,l}+2 \mathfrak{R}\{\mathbf{\bar{h}}_{c,l}^H\tilde{\mathbf{T}}_l\mathbf{e}_{c,l}\}
+ \mathbf{\bar{h}}_{c,l}^{H}\tilde{\mathbf{T}}_l\mathbf{\bar{h}}_{c,l} \geq (2^{\bar{R}_{c,l}} \!-\! t_1)(\sigma_{c,l}^{2}\!+\!\frac{\sigma_{p,l}^{2}}{\rho_{c,l}}).
\end{equation}
In addition, we introduce $ a_l = \frac{1}{\rho_{c,l}} $ and $ b_l = \frac{1}{1-\rho_{c,l}}$  to convert the non-convex constraints into convex ones. According to Lemma 1, by using a slack variable $\lambda_c$, \eqref{eq:Robust_sec_rate_constraint_modified zhu} can be expressed as
\begin{equation} \label{eq:Two_LMI_constraints_user_and_user_energy_modified01}
\!\!\!\!  \left[\begin{array}{cc}
\lambda_{c,l}\mathbf{I} \!+\! \tilde{\mathbf{T}}_l \!&\! \tilde{\mathbf{T}}_l\mathbf{\bar{h}}_{c,l}\\
\mathbf{\bar{h}}_{c,l}^{H}\tilde{\mathbf{T}}_l \!&\! \mathbf{\bar{h}}_{c,l}^{H}\tilde{\mathbf{T}}_l\mathbf{\bar{h}}_{c,l} \!-\! (2^{\bar{R}_{c,l}} \!-\! t_1)(\sigma_{c,l}^{2}\!+\! a_l \sigma_{p,l}^{2}) \!-\! \lambda_{c,l}\varepsilon_{c,l}^{2}
\end{array}
\right] \succeq \mathbf{0}.
\end{equation}

Let us define $ \mathbf{\bar{h}}_{e,k} \triangleq \emph{vec}(\mathbf{H}_{e,k})$.
Using Lemma 1 again and the property $\emph{vec}(\mathbf{M}_1\mathbf{M}_2\mathbf{M}_3) = (\mathbf{M}_3^T \otimes \mathbf{M}_1 )\emph{vec}(\mathbf{M}_2)$ \cite{Matrix_Cookbook}, constraints
  \eqref{eq:Robust_harvested_energy_constraint_user} and \eqref{eq:Robust_harvested_energy_constraint_eve} become
 \begin{subequations}
 \begin{eqnarray}
 &&\!\!\!\!\!\!\!\!\!\!\!\!\!\!\!\!\!\!\!\!\!\!\!\!\!\!   \left[\begin{array}{cc}
\alpha_{c,l}\mathbf{I} \!+\! \mathbf{Q}_{W} \!&\! \mathbf{Q}_{W}\mathbf{\bar{h}}_{c,l}\\
\mathbf{\bar{h}}_{c,l}^{H}\mathbf{Q}_{W} \!&\! \mathbf{\bar{h}}_{c,l}^{H}\mathbf{Q}_{W}\mathbf{\bar{h}}_{c,l} \!+\! \sigma_{c,l}^{2} \!-\! b_l \hat{E}_{c,l} \!-\! \alpha_{c,l}\varepsilon_{c,l}^{2}
\end{array}
\right] \succeq \mathbf{0}, \label{eq:Two_LMI_constraints_user_and_user_energy_modified02}\\
 &&\!\!\!\!\!\!\!\!\!\!\!\!\!\!\!\!\!\!\!\!\!\!\!\!\!\!   \left[\begin{array}{cc}
\alpha_{e,k}\mathbf{I}\!+\!\big(\mathbf{I}\otimes\mathbf{Q}_{W}\big) \!&\!
\big(\mathbf{I}\otimes\mathbf{Q}_{W}\big)\mathbf{\bar{h}}_{e,k} \\
\mathbf{\bar{h}}_{e,k}^{H}\big(\mathbf{I}\otimes\mathbf{Q}_{W}\big)
\!&\! \theta_{e,l} \end{array}\right] \!\succeq\! \mathbf{0}, \forall k, \label{eq:Robust_masked_beamforming_LMI_eve_energy}
\end{eqnarray}
\end{subequations}
where $\alpha_{c,l}$ and $\alpha_{e,k}$ are slack variables, and $\mathbf{Q}_{W}=\mathbf{Q} \!+\! \mathbf{W}$, and $\theta_{e,l} = \mathbf{\bar{h}}_{e,k}^{H}\big(\mathbf{I}\otimes(\mathbf{Q}\!+\!\mathbf{W})\big)\mathbf{\bar{h}}_{e,k} \!-\!  \hat{E}_{e,k} \!+\! N_{R} \sigma_{k}^{2} \!-\! \alpha_{e,k}\varepsilon_{e,k}^{2}$.
To transform the constraint \eqref{11b} into a tractable convex LMI, we exploit the following lemma.

\emph{\underline{Lemma 2 \cite{15LiQ_13TSP_Spatially_selective_AN}}:} For $\mathbf{F}_1, \mathbf{F}_2, \mathbf{F}_3 \in \mathbb{C}^{M\times M}$, we denote $ g(\mathbf{Z}) = \mathbf{Z}^{H}\mathbf{F}_1\mathbf{Z} +
\mathbf{Z}^{H}\mathbf{F}_2 + \mathbf{F}_2^{H}\mathbf{Z} + \mathbf{F}_3 $, satisfying $g(\mathbf{Z}) \succeq \mathbf{0}$ for $
\forall \mathbf{Z} \in \big\{ \mathbf{Z}|\textrm{tr}(\mathbf{Z}^{H}\mathbf{F}_4\mathbf{Z}) \leq 1 \big\}$ with $\mathbf{F}_4 \succeq \mathbf{0}$. Then, the following LMI holds:
\begin{eqnarray}
 \left[\begin{array}{cc}
\mathbf{F}_3 & \mathbf{F}_2^{H}\\
\mathbf{F}_2 & \mathbf{F}_1
\end{array}
\right] - \alpha \left[\begin{array}{cc}
\mathbf{I} & \mathbf{0}   \\
\mathbf{0} & -\mathbf{F}_4
\end{array}
\right] \succeq \mathbf{0}, \nonumber
\end{eqnarray}
where $ \alpha \geq 0 $. ~~~~~~~~~~~~~~~~~~~~~~~~~~~~~~~~~~~~~~~~~~~~~~~~~~~~~~$\blacksquare $

By Lemma 2, the constraint \eqref{11b} can be equivalently given as
\begin{eqnarray} \label{eq:Robust_masked_beamforming_LMI_eve}
\!\!\!\!\!\! \left[\!\!\begin{array}{cc}
\big(({\textstyle{1 \over t_1}} \!-\! 1)\sigma_{k}^{2} \!-\! \lambda_{e,k}\big)\mathbf{I} \!+\!
\mathbf{\bar{H}}_{e,k}^{H}\mathbf{W}_Q\mathbf{\bar{H}}_{e,k}
 \!&\! \mathbf{\bar{H}}_{e,k}^{H}\mathbf{W}_Q \\
\mathbf{W}_Q\mathbf{\bar{H}}_{e,k} \!&\! \mathbf{W}_Q \!+\! \frac{ \lambda_{e,k}}{\varepsilon_{e,k}^{2}}\mathbf{I}
\end{array}
\!\!\right] \succeq \mathbf{0},
\end{eqnarray}
where $\lambda_{e,k}$ is a slack variable and $\mathbf{W}_Q = \big(({\textstyle{1 \over t_1}} \!-\! 1)\mathbf{W} \!-\! \mathbf{Q}\big)$.
According to \eqref{eq:Robust_sec_rate_constraint_modified}-\eqref{eq:Robust_masked_beamforming_LMI_eve}, the WCR-PM problem is now given as
\begin{equation} \label{eq:Masked_beamforming_robust_sec_rate_opt_results}
 \begin{split}
&\min_{\mathbf{Q},{\kern 1pt}\mathbf{W},{\kern 1pt}t_1,{\kern 1pt} a_l,{\kern 1pt} b_l }   ~~ \textrm{tr}(\mathbf{Q}) \\
&~~ \mbox{s.t.}~~    \textrm{tr}(\mathbf{Q}+\mathbf{W}) \leq P, \frac{1}{a_l}+\frac{1}{b_l}\leq1, \textrm{rank}(\mathbf{Q}) = 1, \\
& ~~~~~~~~  \mathbf{Q} \!\succeq\! \mathbf{0}, \mathbf{W} \succeq \mathbf{0}, \lambda_{c,l} \geq 0, \lambda_{e,k} \geq 0,\\
& ~~~~~~~~  \alpha_{c,l} \geq 0, \alpha_{e,k} \geq 0,  a_l \geq 1, b_l \geq 1,\forall l, \forall k,\\
& ~~~~~~~~  \eqref{eq:Two_LMI_constraints_user_and_user_energy_modified01}, \eqref{eq:Two_LMI_constraints_user_and_user_energy_modified02}, \eqref{eq:Robust_masked_beamforming_LMI_eve_energy},\eqref{eq:Robust_masked_beamforming_LMI_eve}.
\end{split}
\end{equation}
By removing the nonconvex constraint $ \textrm{rank}(\mathbf{Q}) = 1 $, the above problem \eqref{eq:Masked_beamforming_robust_sec_rate_opt_results}
becomes convex and can be solved by applying a solver in \cite{28CVX} given $ t_1 $.
Tightness of the relaxation of
\eqref{eq:Masked_beamforming_robust_sec_rate_max_problem_ori} is shown by the following theorem.

\emph{\underline{Theorem 3:}} Provided that the robust problem
\eqref{eq:Masked_beamforming_robust_sec_rate_max_problem_ori} is feasible for a
given $ t_1 $, there always exists an optimal solution $ \mathbf{Q} $  with $ \textrm{rank}(\mathbf{Q}) = 1 $.

\emph{\underline{Proof:}} See Appendix B. ~~~~~~~~~~~~~~~~~~~~~~~~~~~~~~~~~~~~~~~~~~~~~~~~~~~~$\blacksquare $

So far, we have tackled the WCR-PM problem \eqref{eq:Masked_beamforming_robust_sec_rate_opt_results} by deriving a tight rank relation in Theorem 3. Note that problem \eqref{eq:Masked_beamforming_robust_sec_rate_max_problem_ori} can also be solved by applying
one-dimensional line search over $t_1$ as in Section III.
\subsection{Low-Complexity SPCA  Algorithm}
Now, let us consider another reformulation of the WCR-PM problem \eqref{eq:Masked_beamforming_robust_sec_rate_max_problem_ori}
based on the SPCA algorithm.
The optimization framework can also be recast as a convex form by incorporating
channel uncertainties. First, the robust secrecy rate  \eqref{eq:Robust_sec_rate_constraint} can be rewritten as
\begin{subequations}\label{eq:99}
\begin{eqnarray}
&&\!\!\!\!\!\!\!\!\!\!\!\!\!\!\!\!\!\!\!\!\!\!  \log( r_3r_4 ) \geq  \bar{R}_{c,l}, \forall l, \label{eq:99a}\\
&&\!\!\!\!\!\!\!\!\!\!\!\!\!\!\!\!\!\!\!\!\!\!  1 \!+\! \frac{(\mathbf{\bar{h}}_{c,l}\!+\!\mathbf{e}_{c,l})^{H}\mathbf{Q}(\mathbf{\bar{h}}_{c,l}\!+\!\mathbf{e}_{c,l})}{\sigma_{c,l}^{2}\!+\!
(\mathbf{\bar{h}}_{c,l}\!+\!\mathbf{e}_{c,l})^{H}\mathbf{W}(\mathbf{\bar{h}}_{c,l}\!+\!\mathbf{e}_{c,l})\!+\!\frac{\sigma_{p,l}^{2}}{\rho_{c,l}}}  \!\geq\! r_3, \label{eq:99b}\\
&&\!\!\!\!\!\!\!\!\!\!\!\!\!\!\!\!\!\!\!\!\!\!  1 + \frac{{\rm{tr}}\left({({{ {\mathbf{{\bar{H}}}}}}_{e,k}+\mathbf{E}_{e,k})^H
 {{{\mathbf{{Q}}}}}({{ {\mathbf{{\bar{H}}}}}}_{e,k}+\mathbf{E}_{e,k})}\right)}
 {{\sigma _k^2 + {\rm{tr}}\left(({{ {\mathbf{\bar{H}}}}}_{e,k}+\mathbf{E}_{e,k})^H{\mathbf{{W}}}({{ {\mathbf{\bar{H}}}}}_{e,k}+\mathbf{E}_{e,k})\right)}}  \leq \frac{1}{r_4}, \label{eq:99c}
\end{eqnarray}
\end{subequations}
where $r_3 > 0$ and $r_4 > 0$ are slack variables.
The inequalities in \eqref{eq:99} can be rearranged, which gives
\begin{subequations}
\begin{eqnarray}
&&\!\!\!\!\!\!\!\!\!\!\!\!\!\!\!\!\!\!\!\!   r_3r_4  \geq  2^{\bar{R}_{c,l}}, \forall l, \label{eq:25a}\\
&&\!\!\!\!\!\!\!\!\!\!\!\!\!\!\!\!\!\!\!\! \sigma_{c,l}^{2}\!+\!
\mathbf{w}^H(\mathbf{{H}}_{c,l}\!+\!\Delta_{c,l})\mathbf{w} +\frac{\sigma_{p,l}^{2}}{\rho_{c,l}}\leq \frac{\mathbf{q}^H(\mathbf{{H}}_{c,l}\!+\!\Delta_{c,l})\mathbf{q}}{{r_3} -1}, \label{eq:25b}\\
&&\!\!\!\!\!\!\!\!\!\!\!\!\!\!\!\!\!\!\!\!  {{{\sigma _k^2}}+ \mathbf{w}^{H}(\mathbf{\hat{H}}_{e,k}\!+\!\Delta_{e,k})\mathbf{w} +\mathbf{q}^{H}(\mathbf{\hat{H}}_{e,k}\!+\!\Delta_{e,k})\mathbf{q} } \nonumber \\
&&~~~~~~~~~~~~~~~~ \leq \frac{{{\sigma _k^2 + \mathbf{w}^{H}(\mathbf{\hat{H}}_{e,k}\!+\!\Delta_{e,k})\mathbf{w}}}}{r_2},  \label{eq:25c}
\end{eqnarray}
\end{subequations}
where $\Delta_{c,l} = \mathbf{\bar{h}}_{c,l}\mathbf{e}_{c,l}^H+\mathbf{e}_{c,l}\mathbf{\bar{h}}_{c,l}^H+\mathbf{e}_{c,l}\mathbf{e}_{c,l}^H$ and $\Delta_{e,k} = \mathbf{\bar{H}}_{e,k}\mathbf{E}_{e,k}^H+\mathbf{E}_{e,k}\mathbf{\bar{H}}_{e,k}^H+\mathbf{E}_{e,k}\mathbf{E}_{e,k}^H$ stand for
the CSI uncertainty.
It is straightforward to show that
\begin{equation}
\begin{split}
\displaystyle  \left\| {{\Delta _{c,l}}} \right\|_F & \le \| \mathbf{\bar{h}}_{c,l}\mathbf{e}_{c,l}^H \|_F{\rm{ + }}\| \mathbf{e}_{c,l}\mathbf{\bar{h}}_{c,l}^H \|_F{\rm{ + }}\| \mathbf{e}_{c,l}\mathbf{e}_{c,l}^H \|_F\\
 \displaystyle & \le  {\| \mathbf{\bar{h}}_{c,l} \|}   \| \mathbf{e}_{c,l}^H  \| + \|\mathbf{e}_{c,l}\| \| \mathbf{\bar{h}}_{c,l}^H \|{\rm{ + }}{\| \mathbf{e}_{c,l} \|^2}\\
 \displaystyle  &  =  \varepsilon_{c,l}^2 + 2\varepsilon_{c,l}\|\mathbf{\bar{h}}_{c,l}\|,
\end{split}
\end{equation}
\begin{equation}
\begin{split}
\displaystyle  \left\| {{\Delta _{e,k}}} \right\|_F  & \le \| \mathbf{\bar{H}}_{e,k}\mathbf{E}_{e,k}^H \|_F{\rm{ + }}\| \mathbf{E}_{e,k}\mathbf{\bar{H}}_{e,k}^H \|_F{\rm{ + }}\| \mathbf{E}_{e,k}\mathbf{E}_{e,k}^H \|_F\\
 \displaystyle & \le  {\| \mathbf{\bar{H}}_{e,k} \|_F}   \| \mathbf{E}_{e,k}^H  \|_F + \|\mathbf{E}_{e,k}\| _F\| \mathbf{\bar{H}}_{e,k}^H \|_F{\rm{ + }}{\| \mathbf{E}_{e,k} \|_F^2}\\
 \displaystyle  &  =  \varepsilon_{e,k}^2 + 2\varepsilon_{e,k}\|\mathbf{\bar{H}}_{e,k}\|_F.
\end{split}
\end{equation}
Note that ${{\Delta _{c,l}}}$ and  ${{\Delta _{e,k}}}$  are  norm-bounded matrices as $\left\| {{\Delta _{c,l}}} \right\|_F \leq \xi_{c,l}$ and $\left\| {{\Delta _{e,k}}} \right\|_F \leq \xi_{e,k}$ where $\xi_{c,l} = \varepsilon_{c,l}^2 + 2\varepsilon_{c,l}\|\mathbf{\bar{h}}_{c,l}\|$ and $\xi_{e,k} = \varepsilon_{e,k}^2 + 2\varepsilon_{e,k}\|\mathbf{\bar{H}}_{e,k}\|_F$. Similarly, we equivalently recast \eqref{eq:25a} as
\begin{equation} \label{eq:26}
\left\|\left[\sqrt{2^{\bar{R}_{c,l}+2}} ~~~  r_3-r_4\right]\right\|\leq r_3+r_4, ~\forall l.
\end{equation}
~~According to \cite{9zhu16_JCN}, we can minimize constraint \eqref{eq:99b} by maximizing the LHS of \eqref{eq:25b} while minimizing its the RHS. Then the constraints \eqref{eq:25b} and \eqref{eq:25c} can be approximately rewritten as, respectively,
\begin{equation} \label{eq:zzy10}
\begin{split}
&\!\!\!\!\!\!\max_{\|\Delta_{c,l}\|_F\leq  \xi_{c,l}} ~  \sigma_{c,l}^{2}\!+\!
\mathbf{w}^H\mathbf{{A}}_{c,l}\mathbf{w} +\frac{\sigma_{p,l}^{2}}{\rho_{c,l}} \\
&\!\!\!\!\!\! \leq \min_{\|\Delta_{c,l}\|_F\leq  \xi_{c,l}} ~  \frac{\mathbf{q}^H\mathbf{{A}}_{c,l}\mathbf{q}}{{r_3} -1},
\end{split}
\end{equation}
\begin{equation} \label{eq:zzy11}
\begin{split}
&\!\!\!\!\!\! \max_{\|\Delta_{e,k}\|_F\leq  \xi_{e,k}} ~{{{\sigma _k^2}}+ \mathbf{w}^{H}\mathbf{{B}}_{e,k}\mathbf{w} +\mathbf{q}^{H}\mathbf{{B}}_{e,k}\mathbf{q} } \\
&\!\!\!\!\!\! \leq  \min_{\|\Delta_{e,k}\|_F\leq  \xi_{e,k}} \frac{{{\sigma _k^2 + \mathbf{w}^{H}\mathbf{{B}}_{e,k}\mathbf{w}}}}{r_2},
\end{split}
\end{equation}
where $\mathbf{{A}}_{c,l} = \mathbf{{H}}_{c,l}\!+\!\Delta_{c,l}$ and $\mathbf{{B}}_{e,k} = \mathbf{\hat{H}}_{e,k}\!+\!\Delta_{e,k}$.

In order to minimize the RHS of \eqref{eq:zzy10} and \eqref{eq:zzy11}, a loose approximation \cite{loose_approxi} is applied, which gives
\begin{equation} \label{eq:zzy12}
\begin{split}
&\min_{\|\Delta_{c,l}\|_F\leq  \xi_{c,l}} ~
\frac{\mathbf{q}^H\mathbf{{A}}_{c,l}\mathbf{q}}{{r_3} -1} \geq \frac{\mathbf{q}^H\mathbf{{\check{H}}}_{c,l}\mathbf{q}}{{r_3} -1},\\
& \min_{\|\Delta_{e,k}\|_F\leq  \xi_{e,k}} \frac{{{\sigma _k^2 + \mathbf{w}^{H}\mathbf{{B}}_{e,k}\mathbf{w}}}}{r_2} \geq \frac{{{\sigma _k^2 + \mathbf{w}^{H}\mathbf{{\check{H}}}_{e,k}\mathbf{w}}}}{r_2},
\end{split}
\end{equation}
where $\mathbf{{\check{H}}}_{c,l}=\mathbf{{H}}_{c,l}\!-\!\xi_{c,l}\mathbf{I}_{N_T}$ and $\mathbf{{\check{H}}}_{e,k} = \mathbf{\hat{H}}_{e,k}\!-\!\xi_{e,k}\mathbf{I}_{N_T}$.
Using similar technique to the LHS of \eqref{eq:zzy10} and \eqref{eq:zzy11} yields
 \begin{equation}  \label{eq:zzy13}
\!\!\!  \max_{\|\Delta_{c,l}\|_F\leq  \xi_{c,l}} {\kern 1pt}
\mathbf{w}^H\mathbf{{A}}_{c,l}\mathbf{w} \leq \mathbf{w}^H\mathbf{{{\bar{H}}}}_{c,l}\mathbf{w},
\end{equation}
\begin{equation}  \label{eq:zzy14}
\begin{split}
&\max_{\|\Delta_{e,k}\|_F\leq  \xi_{e,k}} ~
{{\sigma _k^2}}+ {\mathbf{w}^{H}\mathbf{{B}}_{e,k}\mathbf{w} +\mathbf{q}^{H}\mathbf{{B}}_{e,k}\mathbf{q} } \\
&\leq {{\sigma _k^2}}+ \mathbf{w}^H\mathbf{{\bar{H}}}_{e,k}\mathbf{w} + \mathbf{q}^H\mathbf{{\bar{H}}}_{e,k}\mathbf{q},
\end{split}
\end{equation}
where $\mathbf{{{\bar{H}}}}_{c,l}=\mathbf{{H}}_{c,l}\!+\!\xi_{c,l}\mathbf{I}_{N_T}$ and $\mathbf{{\bar{H}}}_{e,k} = \mathbf{\hat{H}}_{e,k}\!+\!\xi_{e,k}\mathbf{I}_{N_T}$.

From \eqref{eq:zzy10}-\eqref{eq:zzy14}, \eqref{eq:25b} and \eqref{eq:25c} can be given as
\begin{equation}\label{eq:zzy15}
 \sigma_{c,l}^{2}+ \mathbf{w}^H\mathbf{{\bar{H}}}_{c,l}\mathbf{w} +\frac{\sigma_{p,l}^{2}}{\rho_{c,l}} \leq
\frac{\mathbf{q}^H\mathbf{{\check{H}}}_{c,l}\mathbf{q}}{{r_3} -1},
\end{equation}
\begin{equation}\label{eq:zzy16}
\begin{split}
 {{\sigma _k^2}}+\mathbf{w}^H\mathbf{{\bar{H}}}_{e,k}\mathbf{w} + \mathbf{q}^H\mathbf{{\bar{H}}}_{e,k}\mathbf{q} \leq  \frac{{{\sigma _k^2 + \mathbf{w}^{H}\mathbf{{\check{H}}}_{e,k}\mathbf{w}}}}{r_4}.
\end{split}
\end{equation}
Exploiting the same method in \eqref{eq:zzy1}-\eqref{eq:zzy2}, we obtain
\begin{subequations} \label{eq:zzy17}
\begin{eqnarray}
&&\!\!\!\!\!   \sigma_{c,l}^{2}+ \mathbf{w}^H\mathbf{{\bar{H}}}_{c,l}\mathbf{w} +\frac{\sigma_{p,l}^{2}}{\rho_{c,l}}   \leq F_{\mathbf{{\check{H}}}_{c,l},1}(\mathbf{q},r_3,\mathbf{\tilde{q}},\tilde{r}_3), \label{eq:zzy17a} \\
&&\!\!\!\!\! {{\sigma _k^2}}+\mathbf{w}^H\mathbf{{\bar{H}}}_{e,k}\mathbf{w} + \mathbf{q}^H\mathbf{{\bar{H}}}_{e,k}\mathbf{q}     \nonumber \\
&&\!\!\!\!\!\!\!\!\!\!  \leq \sigma _k^2(\frac{2}{\tilde{r}_4} - \frac{r_4}{\tilde{r}^2_4})+  F_{\mathbf{{\check{H}}}_{e,k},0}(\mathbf{w},r_4,\mathbf{\tilde{w}},\tilde{r}_4). \label{eq:zzy17b}
\end{eqnarray}
\end{subequations}

By using a loose approximation approach for constraints \eqref{eq:Robust_harvested_energy_constraint_user} and \eqref{eq:Robust_harvested_energy_constraint_eve}, we have
\begin{subequations}\label{eq:zzy18}
\begin{eqnarray}
&&\!\!\!\!\!\!\!\!\!\! \mathbf{q}^H\mathbf{{\check{H}}}_{c,l}\mathbf{q} + \mathbf{w}^H\mathbf{{\check{H}}}_{c,l}\mathbf{w}  \!\geq\! \frac{\hat{E}_{c,l}}{1\!-\!\rho_{c,l}} -  \sigma_{c,l}^{2}, \label{eq:zzy18a}\\
&&\!\!\!\!\!\!\!\!\!\! \mathbf{w}^H\mathbf{{\check{H}}}_{e,k}\mathbf{w} + \mathbf{q}^H\mathbf{{\check{H}}}_{e,k}\mathbf{q}  ~ \geq {\hat{E}_{e,k}}  \!-\! N_{R}\sigma_{k}^{2}. \label{eq:zzy18b}
\end{eqnarray}
\end{subequations}
Substituting $\mathbf{q} = {\mathbf{\tilde{q}}}+\Delta \mathbf{q}$ and $\mathbf{w} = {\mathbf{\tilde{w}}}+\Delta \mathbf{w}$ into the LHS of  \eqref{eq:zzy18a} and \eqref{eq:zzy18b}, the SPCA technique can be applied to approximate \eqref{eq:zzy18a} and \eqref{eq:zzy18b}, respectively, as
\begin{subequations}\label{eq:zzy19}
\begin{eqnarray}
&&\!\!\!\!\!\!\!\!\!\! {\mathbf{\tilde{q}}}^H\mathbf{\check{H}}_{c,l}{\mathbf{\tilde{q}}} \!+\! 2\Re \{ {\mathbf{\tilde{q}}}^H\mathbf{\check{H}}_{c,l}\Delta {\mathbf{q}}\} \!+\!
{\mathbf{\tilde{w}}}^H\mathbf{\check{H}}_{c,l}{\mathbf{\tilde{w}}}\nonumber \\
&&\!\!\!\!\!\!\!\!\!\! + 2\Re \{ {\mathbf{\tilde{w}}}^H\mathbf{\check{H}}_{c,l}\Delta {\mathbf{w}}\} \!\geq\! \frac{\hat{E}_{c,l}}{1\!-\!\rho_{c,l}} -  \sigma_{c,l}^{2}, \label{eq:zzy19a}\\
&&\!\!\!\!\!\!\!\!\!\!  {\mathbf{\tilde{q}}}^H\mathbf{\check{H}}_{e,k}{\mathbf{\tilde{q}}} \!+\! 2\Re \{ {\mathbf{\tilde{q}}}^H\mathbf{\check{H}}_{e,k}\Delta {\mathbf{q}}\} \!+\!
{\mathbf{\tilde{w}}}^H\mathbf{\check{H}}_{e,k}{\mathbf{\tilde{w}}} \nonumber \\
&&\!\!\!\!\!\!\!\!\!\! + 2\Re \{ {\mathbf{\tilde{w}}}^H\mathbf{\check{H}}_{e,k}\Delta {\mathbf{w}}\} \geq {\hat{E}_{e,k}}  \!-\! N_{R}\sigma_{k}^{2}. \label{eq:zzy19b}
\end{eqnarray}
\end{subequations}

Eventually, the WCR-PM problem is converted into the following convex form as
\begin{equation}\label{zhu32}
 \begin{split}
& ~~~~ ~~~~  \min_{\mathbf{q},{\kern 1pt}\mathbf{w},{\kern 1pt}\rho_{c,l}, {\kern 1pt}r_3, {\kern 1pt}r_4}  ~~~~ ~ \|\mathbf{q}\| ~~~  \\
& \mbox{s.t.}~ \eqref{eq:26}, {\kern 1pt}\eqref{eq:zzy17a}, {\kern 1pt}\eqref{eq:zzy17b}, {\kern 1pt}\eqref{eq:zzy19a}, {\kern 1pt}\eqref{eq:zzy19b}, {\kern 1pt}\eqref{eq: zzy8}, {\kern 1pt}0 < \rho_{c,l} \leq 1.
\end{split}
\end{equation}
Given $\mathbf{\tilde{q}}$, $\mathbf{\tilde{w}}$, $\tilde{r}_3$, and $\tilde{r}_4$, problem \eqref{zhu32} is convex and can be solved by employing an interior-point method to update iteratively until convergence.
\section{Computational Complexity}
In this section, we evaluate the computational complexity of the proposed robust methods.
As will be shown in Section VI, the proposed SPCA algorithm achieves substantial improvement in complexity for the same performance compared with the method based on 1-D search.
Now we compare complexity of the algorithms through analyses similar to that in \cite{27Complexity_Outage} and \cite{Complexity_zhu}.
The complexity of the proposed algorithms are shown in Table I on the top of next page.
We denote $n$, $D = {{{\log }_2} {\frac{{{t _{\max }} - {t _{\min}}}}{\eta }} } $, and $Q$ as the number of decision variables, the 1-D search size, and the SPCA iteration number, respectively. The complexity analysis is given in the following.

\emph{1)   PM with 1-D Search} in problem \eqref{eq:relaxed_power_min_problem_results} involves $K$ LMI constraints of size $N_{{R}}+1$, two LMI constraints of size $N_{{T}}$,  and  $4L+K+1$ linear constraints. 

\emph{2)  PM with SPCA} in problem \eqref{16} has $L$ SOC constraints of dimension $2$, $L$ SOC constraints of dimension $N_{{T}}+1$, $K$ SOC constraints of dimension $2N_{{T}}+1$, one SOC constraints of dimension $2N_{{T}}$, and $L+3K$ linear constraints. 

\emph{3)  WCR-PM with 1-D Search} in problem \eqref{eq:Masked_beamforming_robust_sec_rate_opt_results} consists of $2L$ LMI constraints of size $N_{{T}}+1$, $K$ LMI constraints of size $N_{{R}}N_{{T}}+1$, two LMI constraints of size $N_{{T}}$, and $5L+2K+1$ linear constraints.

\emph{4)  WCR-PM with SPCA} in problem \eqref{zhu32} contains $L$ SOC constraints of dimension $2$, $L$ SOC constraints of dimension $N_{{T}}+1$, $K$ SOC constraints of dimension $2N_{{T}}+1$, one SOC constraints of dimension $2N_{{T}}$, and $L+3K$ linear constraints.

For example, for a system with $L = 2, K = 3, N_T = 4,  N_R = 2$, $D = 100$, and $Q = 8$, the complexity of the PM with 1-D search, the PM with SPCA, the WCR-PM with 1-D search, and the WCR-PM with SPCA, are ${\cal O}(6.92\times 10^7)$, ${\cal O}(3.70\times 10^5)$, ${\cal O}(7.78\times 10^8)$, and ${\cal O}(1.45\times 10^5)$, respectively. Thus, the complexity of the proposed SPCA method is only 1\% compared to the scheme based on 1-D search.
\begin{table*}[htbp]
\caption{Complexity analysis of the proposed algorithms}
\label{tab:threesome}
\centering
\begin{tabular}{|c|c|}
\hline
Algorithms & Complexity Order   \\
\hline
$\begin{array}{c}  {\textrm{PM with} } \\{\textrm{1-D Search} } \end{array}$ & $\begin{array}{l}  { \mathcal{O}\big( nD\sqrt {KN_R{\rm{+}}2K{\rm{+}}2N_T{\rm{+}}4L{\rm{+}}1}\big\{K(N_R{\rm{+}}1)^3}{ {\rm{+}} 2N_T^3 {\rm{+}}n[K(N_R{\rm{+}}1)^2{\rm{+}}2N_T^2{\rm{+}}4L{\rm{+}}K{\rm{+}}1]{\rm{+}}n^2 \big\} \big)} \\{\textrm{where} ~ n ={\cal O}({2N^2_T} {\rm{+}} L)} \end{array} $ \\
\hline
$\begin{array}{c} {\textrm{PM with} } \\{\textrm{SPCA} } \end{array}$ & $\begin{array}{l}  {  \mathcal{O}\big( nQ\sqrt {5K{\rm{+}}5L{\rm{+}}2}\big\{(2K{\rm{+}}L{\rm{+}}2)N_T{\rm{+}}3L{\rm{+}}K{\rm{+}}}n(3K{\rm{+}}L){\rm{+}}n^2 \big\} \big) ~{\textrm{where} ~ n ={\cal O}({2N_T} {\rm{+}} 2L{\rm{+}}K{\rm{+}}2)} \end{array} $ \\
\hline
$\begin{array}{c}   {\textrm{WCR-PM}} \\ {\textrm{with 1-D}}\\ {\textrm{ Search}}\end{array}$  & $ \begin{array}{l}  {\cal O}\big( n{D}\sqrt {(KN_R{\rm{+}}2L{\rm{+}}2)N_T{\rm{+}}7L{\rm{+}}3K{\rm{+}}1} \big\{2L(N_T{\rm{+}} 1)^3{\rm{+}}K(N_RN_T{\rm{+}}1)^3{\rm{+}}2N_T^3{\rm{+}}n[2L(N_T{\rm{+}}1)^2 \\{\rm{+}} K(N_RN_T{\rm{+}}1)^2{\rm{+}}2N_T^2{\rm{+}}5L{\rm{+}}2K{\rm{+}}1]{\rm{+}}n^2\big\}\big) ~{\textrm{where} ~ n ={\cal O}(2N^2_T {\rm{+}} 4L  {\rm{+}} K)} \end{array} $\\
\hline
$\begin{array}{c} {\textrm{WCR-PM} } \\{\textrm{with SPCA} } \end{array}$ & $\begin{array}{l}  {\mathcal{O}\big( nQ\sqrt {5K{\rm{+}}5L{\rm{+}}2}\big\{(2K{\rm{+}}L{\rm{+}}2)N_T{\rm{+}}3L{\rm{+}}K{\rm{+}}}n(3K{\rm{+}}L){\rm{+}}n^2 \big\} \big) ~{\textrm{where} ~ n ={\cal O}({2N_T} {\rm{+}} L{\rm{+}}2)} \end{array} $ \\
\hline
\end{tabular}
\end{table*}

\section{Numerical Results}\label{six}
In this section, we present numerical results to validate performance of the proposed transmit beamforming schemes.
 In the simulations, we consider a system where the transmitter is equipped with $ N_{T} = 4 $ transmit antennas,
 two CRs are only equipped with single antenna, and three ERs have $ N_{R} = 2$ receive antennas.
Both large-scale and small-scale fading are considered in the channel model. The simplified large-scale fading model is given by $D_{L} = \big(\frac{d}{d_0}\big)^{-\alpha},$
 where $d$ represents the distance between the transmitter and the receiver, $d_0$ is a reference distance equal to $10$ m in this work, and $\alpha= 3$ is the path loss exponent \cite{29fading_channel_model}. We define $d_{c} = 40$ m as the distance between the transmitter and the CRs, and $d_e  = 20$ m as the distance between the transmitter and the ERs, unless otherwise specified.

Because all the receivers are are expected to harvest energy from the RF signal, we consider line-of-sight (LOS) communication scenario where the Rician fading model is adopted for small scale fading coefficients. 
The channel vector $\mathbf{h}_{c,l}$ is expressed as $\mathbf{h}_{c,l} = \sqrt{\frac{K_R}{1+K_R}}\mathbf{h}_{c,l}^{LOS} + \sqrt{\frac{1}{1+K_R}}\mathbf{h}_{c,l}^{NLOS}$,  where $\mathbf{h}_{c,l}^{LOS}$ indicates the LOS deterministic component with
$\| \mathbf{h}_{c,l}^{LOS} \|_2^2 = D_{L}$, $\mathbf{h}_{c,l}^{NLOS}$ represents the Rayleigh fading component as $\mathbf{h}_{c,l}^{NLOS} \sim \mathcal{CN}(0,D_{L}\mathbf{I})$, and $K_R=3$ is the Rician factor. It is noted that for the LOS component, we use the far-field uniform linear antenna array model \cite{30LOS_channel_model}.
In addition, the noise power at the $ l $-th CR is set to be $ \sigma_{c,l}^{2} = -60 ~\textrm{dBm}$ for information transfer and $ \sigma_{p,l}^{2} = -50 ~\textrm{dBm}$ for power transfer.
The noise power at all the ERs is $ \sigma_{k}^{2} = -50 ~\textrm{dBm}, \forall k$.
The channel error bound for the deterministic model is set to $ \varepsilon_{c,l} = \varepsilon_{e,k} = \varepsilon, \forall l,k$.
Consequently, the channel error covariance matrices are given as $\mathbf{N}_{c,l} =\varepsilon^2 \mathbf{I}_{N_T}$ and $\mathbf{D}_k = \varepsilon^2 \mathbf{I}_{N_TN_R}$. The EH efficiency coefficients are set to $\eta_{c,l} = \eta_{e,k} =$ 0.3.

For the perfect CSI case, we compare the PM with SPCA algorithm and the PM with 1-D search method.
For the case with imperfect CSI, we show the performance of the WCR-PM with SPCA algorithm, the WCR-PM with 1-D search method, the no-AN PM with $\mathbf{W} = \mathbf{0}$, and the non-robust method which computes a solution without considering channel uncertainties.

Fig. 1 illustrates the convergence of the SPCA method with respect to iteration numbers for $ P = 50$ dBm, $ \bar{E}_{c,l} = \bar{E}_{e,k} = E$, $R = 1$ bps/Hz, and $\varepsilon =$ 0.01. It is easily seen from the plots that convergence is achieved for all cases within just 8 iterations.

\begin{figure}[!htbp]
\centering
\includegraphics[scale = 0.55]{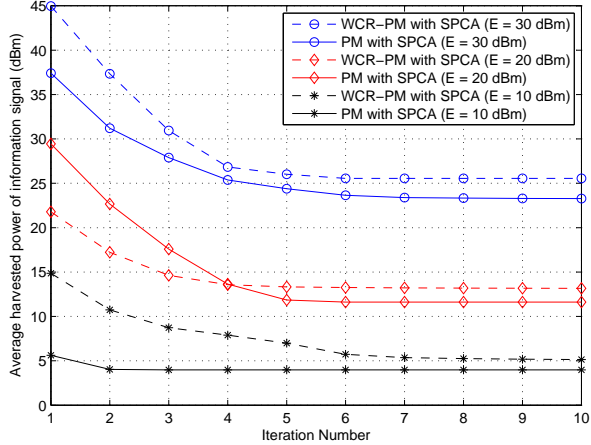}
\caption{Average transmit power of information signal versus iteration numbers}
\label{fig:power_VS_iternumber_161129}
\end{figure}

Fig. 2 illustrates the average transmit power of the information signal in terms of different target secrecy rates with $P = 30$ dBm and  $\bar{E}_{c,l}  = \bar{E}_{e,k} = 10$ dBm, $\forall k$. It is observed that the transmit power increases with the secrecy rate target.
In addition, the SPCA algorithm achieves the same performance as the 1-D search method, but with much lower complexity. Compared with the scheme without AN, the power consumption of the proposed AN-aided scheme is $9$ dB lower. Moreover, we can check that the proposed scheme performs better than the scheme with $\rho_{c} = \rho_{c,l} = 0.5$, and the performance gap becomes larger as the target secrecy rate increases. This indicates that optimizing the PS ratio $\rho_{c,l}$ is important, especially when the target secrecy rate is high.

\begin{figure}[!htbp]
\centering
\includegraphics[scale = 0.55]{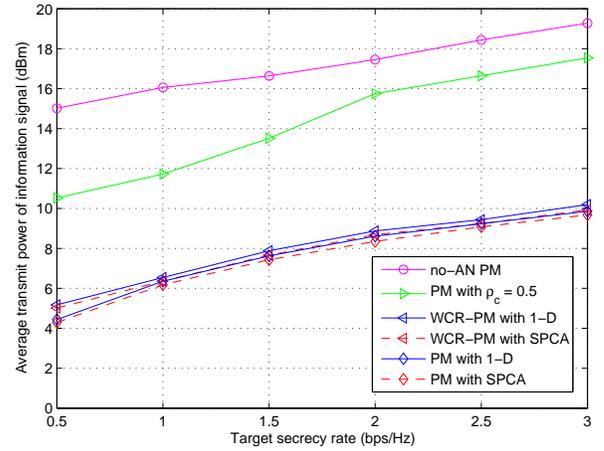}
\caption{Average transmit power of information signal versus target secrecy rate}
\label{fig:power_VS_rate_161129}
\end{figure}



\begin{figure}[!htbp]
\centering
\includegraphics[scale = 0.55]{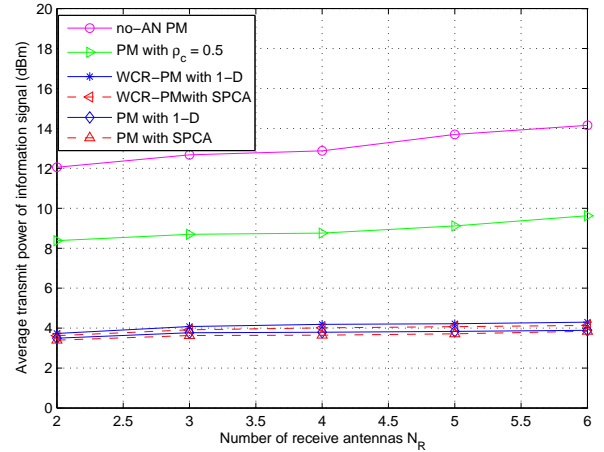}
\caption{Average transmit power of information signal versus the number of receive antenna at the ERs}
\label{fig:power_VS_number_ERantenna_161129}
\end{figure}

In Fig. 3, we compare the average transmit power with respect to different numbers of ER antennas by fixing $N_T = 8$, $P = 40$ dBm, $\bar{E}_{c,l} = \bar{E}_{e,k} = 10$ dBm, and $\bar{R}_{c,l} = 1$ bps/Hz. In this figure, one can observe that the performance of the 1-D search method and that of the proposed SPCA algorithm remains the same regardless of the value of $N_R$. 
This is due to the fact that all the harvested power at the ERs can be provided by the AN signal. 

\begin{figure}[!htbp]
\centering
\includegraphics[scale = 0.55]{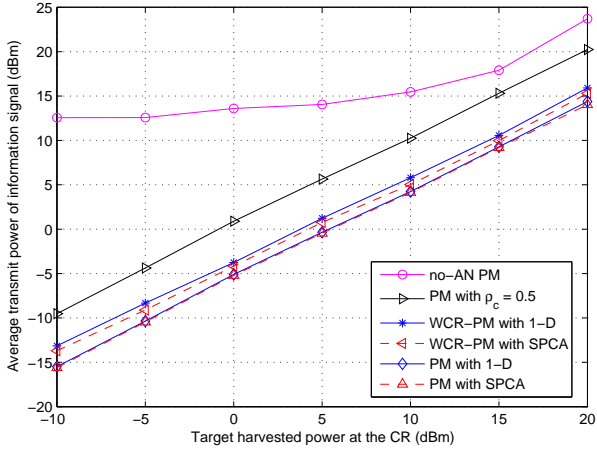}
\caption{Average transmit power of information signal versus the target harvested power at the CR}
\label{fig:power_VS_harvestedpower_161129}
\end{figure}

In Fig. 4, we plot the average transmit power in terms of different target harvested power at the CR with $P = 40$ dBm, $\bar{E}_{e,k} = 10$ dBm and $\bar{R}_{c,l} = 0.5$ bps/Hz. We can check that the curves of the PM and the WCR-PM schemes increase with the same slope.
Moveover, when the harvested power target decreases, the performance gap between the no-AN PM scheme and the proposed PM scheme becomes wider.
This indicates that AN is essential in achieving the performance gains. Furthermore, the PM scheme and the WCR-PM scheme require $6$ dB and $4.5$ dB lower power than the PM scheme with fixed $\rho_c$, respectively.

%
%

\begin{figure}[!htbp]
\centering
\includegraphics[scale = 0.55]{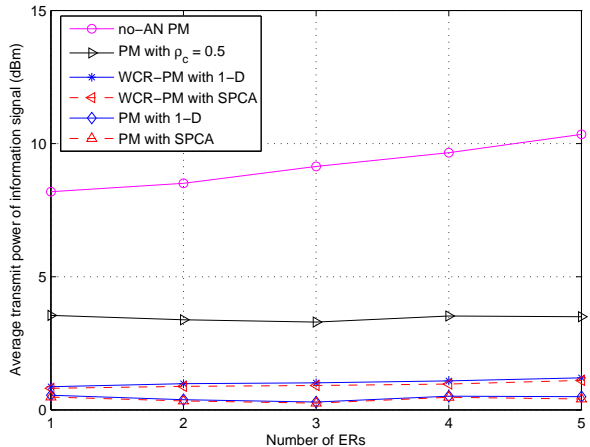}
\caption{Average transmit power of information signal versus the number of ERs}
\label{fig:power_VS_ERnumber_161129}
\end{figure}

Fig. 5 evaluates the average transmit power of information signal with respect to different number of ER with $N_T = 6$, $P = 30$ dBm, $\bar{E}_{c,l} = \bar{E}_{e,k} = 5$ dBm, and $\bar{R}_{c,l} = 1$ bps/Hz. It is observed that both the proposed SPCA algorithms and  the 1-D search method achieved the same performance, and the PM with $\rho_{c} = 0.5$ exhibits a $2.8$ dBm loss over the scheme with the optimal $\rho_{c,l}^*$ regardless of the number of ERs. In addition, we can find that the curve of the no-An PM scheme increases slowly with the number of ERs.


\begin{figure}[!htbp]
\centering
\includegraphics[scale = 0.55]{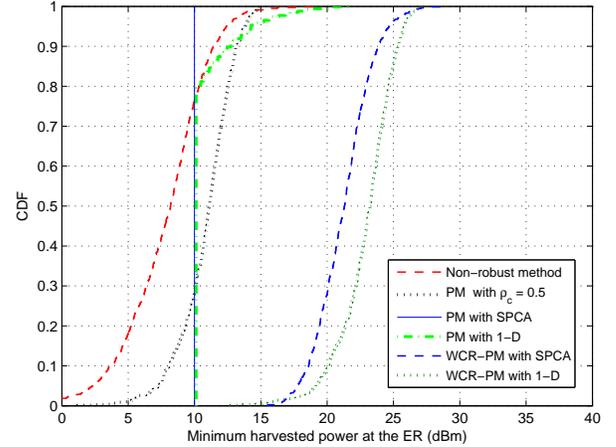}
\caption{CDF of the minimum harvested power at the ERs}
\label{fig:CDF_Harvestedpower_161201_OK}
\end{figure}

Finally, in Fig. 6 we plot the cumulative density function (CDF) of the minimum harvested power at the ERs with $N_T = 4$, $P = 30$ dBm, $\bar{E}_{c,l} = 10$ dBm, $\bar{E}_{e,k} = 10$ dBm, $\bar{R}_{c,l} = 0.5$ bps/Hz, and $\varepsilon =  0.1$.
It is observed that the proposed robust schemes always satisfy the predefined harvested power target ($\bar{E}_{e,k} = 10$ dBm), whereas the non-robust method achieves only 25\% of the predefined harvested power at the ERs.

\section{Conclusion}
In this paper, we have proposed a transmit beamforming power minimization scheme for a multi-user MIMO SWIPT secrecy communication system where power splitters are employed by the receivers for SWIPT operation.
The original problem, which was shown to be non-convex, was relaxed to formulate a two-layer problem.
The inner layer problem was recast as a sequence of SDPs and solved accordingly. Then the optimal solution to the outer problem, on the other hand, has been obtained through one-dimensional line search.
This optimization framework has also been extended to robust secrecy transmission designs by incorporating deterministic channel uncertainties.
Moreover, tightness of the relaxation scheme has been investigated for both the perfect and imperfect CSI cases by showing that the optimal
solution is rank-one.
To reduce the computational complexity, an SPCA based iterative algorithm has been proposed, which achieved near-optimal solution in both the perfect and imperfect CSI cases.
Finally, numerical results have been provided to validate the performance of the proposed transmit beamforming schemes. 

\appendices
\section{Proof of Theorem 2}
We first consider the Lagrange dual function of \eqref{eq:relaxed_power_min_problem_results} as
\begin{eqnarray}
&&\!\!\!\!\!\!\!\!\!\!\!\! \mathcal{L}(\mathbf{Q},\mathbf{W},\mathbf{Z},\mathbf{Y}, \xi_l, \mathbf{A}_{e,k},\gamma, \mu_l, \theta_{k}) \!=\! \textrm{tr}(\mathbf{Q})- \textrm{tr}(\mathbf{Z}\mathbf{Q})  - \textrm{tr}(\mathbf{Y}\mathbf{W}) \nonumber\\
&&\!\!\!\!\!\!\!\!\!\!\!\!- \xi_l\bigg[\textrm{tr}\bigg(\mathbf{h}_{c,l}\mathbf{h}_{c,l}^{H}[t \mathbf{Q} \!-\! (2^{\bar{R}_{c,l}} \!-\! t)\mathbf{W}]\bigg) - (2^{\bar{R}_{c,l}} - t)(\sigma_{c,l}^{2} \!+\! \frac{\sigma_{p,l}^{2}}{\rho_{c,l}})\bigg] \nonumber\\
&&\!\!\!\!\!\!\!\!\!\!\!\!- \sum_{k=1}^{K}\textrm{tr}\bigg[\mathbf{A}_{e,k}\bigg(\mathbf{H}_{e,k}^{H}\big(({\frac{1}{t}}  \!-\! 1)\mathbf{W} \nonumber -\mathbf{Q}\big)\mathbf{H}_{e,k} \!+\! ({\frac{1}{t}} \!-\! 1)\sigma_{k}^{2}\mathbf{I} \bigg)\bigg]  \nonumber  \\
&&\!\!\!\!\!\!\!\!\!\!\!\! + \gamma\big(\textrm{tr}(\mathbf{Q}\!+\!\mathbf{W}) \!-\! P\big)  \!-\! \mu_l\bigg[ \textrm{tr}\big(\mathbf{h}_{c,l}\mathbf{h}_{c,l}^{H}(\mathbf{Q}\!+\!\mathbf{W})\big)
\nonumber  - \frac{\bar{E}_{c,l}}{1\!-\!\rho_{c,l}} \!+\! \sigma_{c,l}^{2} \bigg] \nonumber  \\
&&\!\!\!\!\!\!\!\!\!\!\!\!- \sum_{k=1}^{K} \theta_{k}\bigg( \textrm{tr}\big(\mathbf{H}_{e,k}^{H}(\mathbf{Q}\!+\!\mathbf{W})\mathbf{H}_{e,k}\big)
- \bar{E}_{e,k} \!+\! N_{R}\sigma_{k}^{2} \bigg),\nonumber
\end{eqnarray}
where $ \mathbf{Z} \in \mathbb{H}_{+}^{N_{T}} $, $ \mathbf{Y} \in \mathbb{H}_{+}^{N_{T}} $, $ \xi_l \in \mathbb{R}_{+} $,
$ \mathbf{A}_{e,k} \in \mathbb{H}_{+}^{N_{T}} $, $ \gamma \in \mathbb{R}_{+} $,
$ \mu_l\in \mathbb{R}_{+} $, and $ \theta_{k} $ are the dual variables of $ \mathbf{Q} $, $ \mathbf{W} $,
\eqref{eq:objective_function_with_log_relaxed}, \eqref{eq:Eavesdroppers_rate_slack_variable_log_relaxed},
\eqref{eq:Power_constraints}, \eqref{eq:Energy_constraint_user_PS}, and \eqref{eq:Energy_constraint_eve}, respectively. Then, some of the related KKT conditions are listed as
\begin{subequations}
\begin{eqnarray}
&&\!\!\!\!\!\!\!\!\!\!\! \frac{\partial \mathcal{L}}{\partial \mathbf{Q}}   =
\mathbf{I} - \mathbf{Z} - (\xi_l t + \mu_l) \mathbf{h}_{c,l}\mathbf{h}_{c,l}^{H} +
\sum_{k=1}^{K}\mathbf{H}_{e,k}\mathbf{A}_{e,k}\mathbf{H}_{e,k}^{H} \nonumber \\
 &&~~~~ + \gamma \mathbf{I}  - \sum_{k=1}^{K}\theta_{k}\mathbf{H}_{e,k}\mathbf{H}_{e,k}^{H} = \mathbf{0}, \label{eq:Partial_derivation_Qs}\\
&&\!\!\!\!\!\!\!\!\!\!\! \frac{\partial \mathcal{L}}{\partial \mathbf{W}}   \!= \! -\mathbf{Y} \!+\!
[ \xi_l (2^{\bar{R}_{c,l}} \!-\! t) \!-\! \mu_l]\mathbf{h}_{c,l}\mathbf{h}_{c,l}^{H}  + \gamma\mathbf{I}  \!-\! \sum_{k=1}^{K}\theta_{k}\mathbf{H}_{e,k}\mathbf{H}_{e,k}^{H} \nonumber  \\
 &&~~~~ \!-\!
\sum_{k=1}^{K} ({\frac{1}{t}}\!-\!1)\mathbf{H}_{e,k}\mathbf{A}_{e,k}\mathbf{H}_{e,k}^{H}\!=\!
 \mathbf{0},\!\!\!\!\!\!\!\! \label{eq:Partial_derivation_W}\\
&&\!\!\!\!\!\!\!\!\!\!\! \mathbf{Z}\mathbf{Q}  = \mathbf{0}, ~\mathbf{Y} \!\succeq\! \mathbf{0},
~\mathbf{A}_{e,k} \succeq \mathbf{0}, ~ \xi_l \geq 0,~\mu_l\geq 0,~\forall k. \!\!\!\!\!\!\!\!\label{eq:Other_KKTs}
\end{eqnarray}
\end{subequations}

From the Lagrangian function and the KKT conditions, we have
 $ 0 < \rho_{c,l} \leq 1 $  and the KKT condition $ \xi_l > 0 $ and $ \mu_l> 0 $. Now, we will show these conditions via the dual
problem of \eqref{eq:relaxed_power_min_problem_results} as
\begin{eqnarray}\label{eq:Dual_problem_of_masked_beamforming_power_min}
&&  \!\!\!\!\!\!\!\!\!\! \max_{\mathbf{Z},\mathbf{Y},\mathbf{A}_{e,k},\xi_l,\gamma,\mu_l,\theta_{k}} \min_{\mathbf{Q},\mathbf{W},\rho_{c,l}}
 \mathcal{L}(\mathbf{Q},\mathbf{W},\mathbf{Z},\mathbf{Y}, \xi_l, \mathbf{A}_{e,k},\gamma,  \mu_l) \nonumber\\
=&&\!\!\!\!\!\!\!\!\!\!  \max_{\mathbf{Z},\mathbf{Y},\mathbf{A}_{e,k},\xi_l,\gamma,\mu_l,\theta_{k}} \min_{\mathbf{Q},\mathbf{W},\rho_{c,l}}
\bigg[- \textrm{tr}(\mathbf{Z}\mathbf{Q}) \!-\!  \textrm{tr}(\mathbf{Y}\mathbf{W}) \nonumber \\
&&\!\!\!\!\!\!\! +  \frac{\xi_l(2^{\bar{R}_{c,l}}\! -\! t)
\sigma_{p,l}^{2}}{\rho_{c,l}} \!+\! \frac{\mu_l\bar{E}_{c,l}}{1 \!-\!\rho_{c,l}} \!-\! \sum_{k=1}^{K}({\frac{1}{t}}\!-\! 1)\sigma_{e}^{2}\textrm{tr}
(\mathbf{A}_{e,k}) \\
&& \!\!\!\!\!\! + [\xi_l(2^{\bar{R}_{c,l}} \!-\! t) \!-\! \mu_l]\sigma_{c,l}^{2} \!-\! \gamma P  +
\sum_{k=1}^{K}\theta_{k}(\bar{E}_{e,k} \nonumber \! -  N_{R}\sigma_{k}^{2})\bigg].
\end{eqnarray}

Since problem \eqref{eq:relaxed_power_min_problem_results} is convex and satisfies the
Slater's condition, the duality gap between \eqref{eq:relaxed_power_min_problem_results} and
\eqref{eq:Dual_problem_of_masked_beamforming_power_min} is zero, and the strong duality
 holds. Therefore solving problem \eqref{eq:relaxed_power_min_problem_results} is
 equivalent to solving \eqref{eq:Dual_problem_of_masked_beamforming_power_min}.
In addition, the constraint $ 0 < \rho_{c,l} \leq 1 $ can be satisfied as
\begin{eqnarray}
\min_{0 < \rho_{c,l} \leq 1} \frac{\xi_l(2^{\bar{R}_{c,l}} \!-\! t)\sigma_{p,l}^{2}}{\rho_{c,l}} \!+\! \frac{\mu_l\bar{E}_{c,l}}{1\!-\!\rho_{c,l}}. \nonumber
\end{eqnarray}
Also the optimal variable $\rho_{c,l}^*$, and the dual variables
$\xi_l^*, \mu_l^*$ are related by
\begin{eqnarray}
 \rho_{c,l}^* \!=\! \frac{\sqrt{\xi_l^*(2^{\bar{R}_{c,l}} \!-\! t)\sigma_{p,l}^{2}}}{ \sqrt{\xi_l^*(2^{\bar{R}_{c,l}} \!-\! t)\sigma_{p,l}^{2}} \!+\! \sqrt{\mu_l^* \bar{E}_{c,l}}}.\nonumber
\end{eqnarray}

From the above inequality, we will show that $ \xi_l^* > 0 $ and $ \mu_l^* > 0 $ by
contradiction. Suppose that $ \xi_l^* = 0 $ and/or $ \mu_l^* = 0 $. Then there are two
cases (i.e., $ \rho_{c,l}^* = 0 $ or $ 1 $),
 which violate the constraints \eqref{eq:Achieved_sec_rate_constraint} and
 \eqref{eq:Energy_constraint_user_PS}. Thus, it follows $ \xi_l > 0 $ and $ \mu_l> 0 $. Now, subtracting \eqref{eq:Partial_derivation_W} from \eqref{eq:Partial_derivation_Qs} yields
\begin{eqnarray}\label{eq:Substraction_Z_Y}
 \mathbf{Z} \!+\! \xi_l^* 2^{\bar{R}_{c,l}} \mathbf{h}_{c,l}\mathbf{h}_{c,l}^{H} \!=\! \mathbf{I} \!+\! \mathbf{Y} \!+\! {\frac{1}{t}} \sum_{k=1}^{K} \mathbf{H}_{e,k}\mathbf{A}_{e,k}\mathbf{H}_{e,k}^{H}.
\end{eqnarray}
We post-multiply $ \mathbf{Q} $ by both sides of \eqref{eq:Substraction_Z_Y} and use \eqref{eq:Other_KKTs} as
\begin{eqnarray}
\xi_l^* 2^{\bar{R}_{c,l}}\mathbf{h}_{c,l}\mathbf{h}_{c,l}^{H}\mathbf{Q} \!=\! \bigg(\! \mathbf{I} \!+\! \mathbf{Y} \!+\! {\frac{1}{t}} \sum_{k=1}^{K} \mathbf{H}_{e,k}\mathbf{A}_{e,k}\mathbf{H}_{e,k}^{H} \!\bigg)\mathbf{Q}. \nonumber
\end{eqnarray}

Then, it becomes
\begin{eqnarray}
\xi_l^* 2^{\bar{R}_{c,l}}\bigg( \mathbf{I} \!+\! \mathbf{Y} \!+\! {\frac{1}{t}} \sum_{k=1}^{K} \mathbf{H}_{e,k}\mathbf{A}_{e,k}\mathbf{H}_{e,k}^{H} \bigg)^{-1} \mathbf{h}_{c,l}\mathbf{h}_{c,l}^{H}\mathbf{Q} \!=\! \mathbf{Q}.\!\!\!\!\!\!\!\!\!\!\!\!\!\!\!\!\!\!\nonumber
\end{eqnarray}
Due to $ \xi_l^* > 0 $, we have
\begin{equation*}
\begin{split}
 &\textrm{rank}(\mathbf{Q})\nonumber \\
=& \textrm{rank}\bigg[ \xi_l^* 2^{\bar{R}_{c,l}} \bigg(\! \mathbf{I} \!+\! \mathbf{Y} \!+\! {\frac{1}{t}} \sum_{k=1}^{K} \mathbf{H}_{e,k}\mathbf{A}_{e,k}\mathbf{H}_{e,k}^{H} \!\bigg)^{-1} \mathbf{h}_{c,l}\mathbf{h}_{c,l}^{H}\mathbf{Q} \bigg]  \nonumber \\
=& \textrm{rank}(\mathbf{h}_{c,l}\mathbf{h}_{c,l}^{H}) \leq\! 1.
\end{split}
\end{equation*}
This completes the theorem.~~~~~~~~~~~~~~~~~~~~~~~~~~~~~~~~~~~$ \blacksquare $

\section{Proof of Theorem 3}
First we write the Lagrange dual function of \eqref{eq:Masked_beamforming_robust_sec_rate_opt_results} as
\begin{equation}
\begin{split}
& \mathcal{L}(\mathbf{Q},\mathbf{W},\mathbf{Z},\mathbf{Y},\xi_l,  \mathbf{T}_{c,l},\mathbf{T}_{e,k},\mathbf{R}_{c,l},\mathbf{R}_{e,k}) = \textrm{tr}(\mathbf{Q}) - \textrm{tr}(\mathbf{Z}\mathbf{Q}) \\
& \!-\! \textrm{tr}(\mathbf{Y}\mathbf{W}) \!+\! \xi_l\big(\textrm{tr}(\mathbf{Q}+\mathbf{W}) \!-\! P\big)  - \textrm{tr}(\mathbf{T}_{c,l}\mathbf{A}_{c,l}) \\
&   \!-\! \textrm{tr}\bigg(\mathbf{T}_{c,l}\mathbf{H}_{c,l}^{H}\big(t_1\mathbf{Q} \!-\! (2^{\bar{R}_{c,l}} \!-\! t_1)\mathbf{W}\big)\mathbf{H}_{c,l}\bigg) - \sum_{k=1}^{K}\textrm{tr}(\mathbf{T}_{e,k}\mathbf{A}_{e,k}) \\
&   \!-\! \sum_{k=1}^{K}\textrm{tr}\bigg(\mathbf{T}_{e,k}\mathbf{G}_{e,k}^{H} \big(({\frac{1}{t_1}} \!-\! 1)\mathbf{W}  \!-
 \mathbf{Q}\big) \mathbf{G}_{e,k}\bigg) \!-\! \textrm{tr}(\mathbf{R}_{c,l}\mathbf{B}_{c,l})  \nonumber \\
& \!-\! \textrm{tr}\big(\mathbf{R}_{c,l}\mathbf{H}_{c,l}^{H}(\mathbf{Q} \!+\! \mathbf{W})\mathbf{H}_{c,l}\big)\!-\! \sum_{k=1}^{K}\sum_{j=1}^{N_{R}}\textrm{tr}(\mathbf{R}_{e,k}\mathbf{B}_{e,k})\\
& \!-\sum_{k=1}^{K}\sum_{j=1}^{N_{R}}\textrm{tr}\bigg((\mathbf{Q} \!+\! \mathbf{W})\mathbf{D}_{e,k}^{(j,j)} \bigg), \!\!\!\!\!\!\!
\end{split}
\end{equation}
where $ \mathbf{Z} \in \mathbb{H}_{+}^{N_{T}} $, $ \mathbf{Y} \in \mathbb{H}_{+}^{N_{T}} $, $ \xi_l \in \mathbb{R}_{+} $, $ \mathbf{T}_{c,l} \in \mathbb{H}_{+}^{N_{T}} $,
 $ \mathbf{T}_{e,k} \in \mathbb{H}_{+}^{N_{T}} $, $ \mathbf{R}_{c,l} \in \mathbb{H}_{+}^{N_{T}} $, and
  $ \mathbf{R}_{e,k} \in \mathbb{H}_{+}^{N_{T}N_{R}} $ are the dual variables of $ \mathbf{Q} $, $ \mathbf{W} $,
   \eqref{eq:Two_LMI_constraints_user_and_user_energy_modified01},  \eqref{eq:Robust_masked_beamforming_LMI_eve}, \eqref{eq:Two_LMI_constraints_user_and_user_energy_modified02}, and \eqref{eq:Robust_masked_beamforming_LMI_eve_energy}, respectively, and
\begin{eqnarray}
&&\!\!\!\!\!\!\!\!\!\!\!\!\! \mathbf{A}_{c,l} \!=\! \left[\!\!\begin{array}{cc}
\lambda_{c,l}\mathbf{I} \!&\! \mathbf{0}_{N_T \times 1}\\
\mathbf{0}^{H}_{1 \times N_T} \!&\! -(2^{\bar{R}_{c,l}} \!-\! t_1)(\sigma_{c,l}^{2}\!+\! a \sigma_{p,l}^{2}) \!-\! \lambda_{c,l}\varepsilon_{c,l}^{2},
\end{array}
\!\!\right], \nonumber\\
&&\!\!\!\!\!\!\!\!\!\!\!\!\! \mathbf{A}_{e,k} \!=\! \left[\begin{array}{cc}
(({\frac{1}{t_1}}-1)\sigma_{k}^{2} \!-\! \lambda_{e,k})\mathbf{I} \!&\! \mathbf{0}_{N_T \times 1}\\
\mathbf{0}^{H}_{1 \times N_T} \!&\! \frac{\lambda_{e,k}}{\varepsilon_{e,k}^{2}}\mathbf{I}
\end{array}
\right]\!\!,\mathbf{G}_{e,k} \!=\! \left[\begin{array}{cc}
\mathbf{\bar{H}}_{e,k} \!&\! \mathbf{I}
\end{array}
\right], \nonumber\\
&&\!\!\!\!\!\!\!\!\!\!\!\!\! \mathbf{B}_{c,l} \!=\! \left[\begin{array}{cc}
\alpha_{c,l}\mathbf{I} \!&\! \mathbf{0}_{N_T \times 1} \\
\mathbf{0}^{H}_{1 \times N_T} \!&\! \sigma_{c,l}^{2} \!-\! b \bar{E}_{c,l} \!-\! \alpha_{c,l}\varepsilon_{c,l}^{2}
\end{array}
\right]\!\!,   \mathbf{{\hat{H}}}_{e,k} \!=\! \left[\!\!\begin{array}{cc}
\mathbf{I} \!&\! \mathbf{\bar{h}}_{e,k}
\end{array}
\!\!\right], \nonumber\\
&&\!\!\!\!\!\!\!\!\!\!\!\!\! \mathbf{B}_{e,k} \!=\! \left[\!\!\begin{array}{cc}
\alpha_{e,k}\mathbf{I} \!&\! \mathbf{0}_{N^2_T \times 1} \\
\mathbf{0}^{H}_{1 \times N^2_T} \!&\! -\bar{E}_{e,k} \!+\! N_{R}\sigma_{k}^{2} \!-\! \alpha_{e,k}\varepsilon_{e,k}^{2}
\end{array}
\!\!\right]\!\!, \mathbf{H}_{c,l} \!=\! \left[\!\!\begin{array}{cc}
\mathbf{I}  \!&\!  \mathbf{\bar{h}}_{c,l}
\end{array}
\!\!\right],\nonumber
\end{eqnarray}
and $\mathbf{D}_{e,k}^{(j,j)} \in \mathbb{H}_{+}^{N_{T}} $ is a block submatrix of
$\mathbf{\hat{H}}_{e,k}\mathbf{R}_{e,k}\mathbf{\hat{H}}_{e,k}^{H} $ as
\begin{equation*}
\mathbf{\hat{H}}_{e,k}\mathbf{R}_{e,k}\mathbf{\hat{H}}_{e,k}^{H} \!=\! \left[\!\!\begin{array}{ccc}
      \mathbf{D}^{(1,1)}_{e,k} \!&\! \cdots \!&\! \mathbf{D}^{(1,N_{R})}_{e,k}  \\
      \vdots \!&\! \ddots \!&\! \vdots\\
      \mathbf{D}^{(N_{R},1)}_{e,k} \!&\! \cdots \!&\! \mathbf{D}^{(N_{R},N_{R})}_{e,k}
    \end{array}\!\!\right], \forall k. \\
\end{equation*}

Now, we consider the following related KKT conditions as
\begin{subequations}
\begin{eqnarray}
&&\!\!\!\!\!\!\!  \frac{\partial \mathcal{L}}{\partial \mathbf{Q}} = \mathbf{I} - \mathbf{Z} + \xi_l \mathbf{I} - t_1\mathbf{H}_{c,l}\mathbf{T}_{c,l}\mathbf{H}_{c,l}^{H} + \sum_{k=1}^{K}\mathbf{G}_{e,k}\mathbf{T}_{e,k}\mathbf{G}_{e,k}^{H}  \nonumber \\
&&\!\!\!\!\!\!\!~~~~~~  - \mathbf{H}_{c,l}\mathbf{R}_{c,l}\mathbf{H}_{c,l}^{H} -  \sum_{k=1}^{K}\sum_{j=1}^{N_{R}}\mathbf{D}_{e,k}^{(j,j)} = \mathbf{0}, \label{eq:KKT_for_derivatives_of_Qs}\\
&&\!\!\!\!\!\!\! \frac{\partial \mathcal{L}}{\partial \mathbf{W}} = -\mathbf{Y} + \xi_l \mathbf{I}  +  (2^{\bar{R}_{c,l}} - t_1)\mathbf{H}_{c,l}\mathbf{T}_{c,l}\mathbf{H}_{c,l}^{H}  \nonumber \\
&&\!\!\!\!\!\!\!~~~~~~ - \sum_{k=1}^{K} ({\frac{1}{t_1}} - 1)\mathbf{G}_{e,k}\mathbf{T}_{e,k}\mathbf{G}_{e,k}^{H} - \mathbf{H}_{c,l}\mathbf{R}_{c,l}\mathbf{H}_{c,l}^{H} \nonumber \\
&&\!\!\!\!\!\!\!~~~~~~ - \sum_{k=1}^{K}\sum_{j=1}^{N_{T}} \mathbf{D}_{e,k}^{(j,j)} = \mathbf{0},\label{eq:KKT_for_derivatives_of_W} \\
&&\!\!\!\!\!\!\!~~ \mathbf{Q}\mathbf{Z} = \mathbf{0},~\mathbf{W} \succeq \mathbf {0}, \\
&&\!\!\!\!\!\!\!~ \big(\mathbf{A}_{c,l} +\mathbf{H}_{c,l}^{H}\big( t_1\mathbf{Q} - (2^{\bar{R}_{c,l}} - t_1)\mathbf{W} \big)\mathbf{H}_{c,l} \big)\mathbf{T}_{c,l} = \mathbf{0}. \label{eq:Another_KKTs}
\end{eqnarray}
\end{subequations}

Subtracting \eqref{eq:KKT_for_derivatives_of_W} from \eqref{eq:KKT_for_derivatives_of_Qs}  generates
\begin{equation*}
\mathbf{I} \!-\! \mathbf{Z} \!+\! \mathbf{Y} \!-\! 2^{\bar{R}_{c,l}}\mathbf{H}_{c,l}\mathbf{T}_{c,l}\mathbf{H}_{c,l}^{H} \!+\! \sum_{k=1}^{K} {\frac{1}{t_1}}\mathbf{G}_{e,k}\mathbf{T}_{e,k}\mathbf{G}_{e,k}^{H} \!=\! \mathbf{0}, \nonumber
\end{equation*}
and it follows
\begin{equation*}
  \mathbf{Z} \!+\!  2^{\bar{R}_{c,l}}\mathbf{H}_{c,l}\mathbf{T}_{c,l}\mathbf{H}_{c,l}^{H}   \!=\! \mathbf{I}+\mathbf{Y} + \sum_{k=1}^{K} {\frac{1}{t_1}}\mathbf{G}_{e,k}\mathbf{T}_{e,k}\mathbf{G}_{e,k}^{H}.
\end{equation*}
Pre-multiplying both sides of the above equality by $ \mathbf{Q} $ and applying the inverse yields
\begin{equation*}
2^{\bar{R}_{c,l}}\mathbf{Q}\mathbf{H}_{c,l}\mathbf{T}_{c,l}\mathbf{H}_{c,l}^{H}\bigg(\mathbf{I} \!+\! \mathbf{Y} \!+\! \sum_{k=1}^{K} {\frac{1}{t_1}}\mathbf{G}_{e,k}\mathbf{T}_{e,k}\mathbf{G}_{e,k}^{H}\bigg)^{-1} \!\!\!=\! \mathbf{Q}.
\end{equation*}
Hence, we have the rank relation as
\begin{equation}
\begin{split}
  &\textrm{rank}(\mathbf{Q}) \\
 = ~&\textrm{rank}\bigg( 2^{\bar{R}_{c,l}}\mathbf{Q}\mathbf{H}_{c,l}\mathbf{T}_{c,l}\mathbf{H}_{c,l}^{H}\big(\mathbf{I} \!+\! \mathbf{Y} \!+\! \sum_{k=1}^{K} {\frac{1}{t_1}}\mathbf{G}_{e,k}\mathbf{T}_{e,k}\mathbf{G}_{e,k}^{H}\big)^{-1} \bigg)\\
 \leq ~& \textrm{rank}(\mathbf{H}_{c,l}\mathbf{T}_{c,l}\mathbf{H}_{c,l}^{H}).
\end{split}
\end{equation}

Now, we will compute the rank of the matrix $ \mathbf{H}_{c,l}\mathbf{T}_{c,l}\mathbf{H}_{c,l}^{H} $. To this end, we consider the following two equalities as
\begin{eqnarray}
&& \left[\!\!\begin{array}{cc}
 \mathbf{I} \!&\! \mathbf{0}
\end{array}
\!\!\right] \mathbf{H}_{c,l}^{H} \!=\! \mathbf{I}, \nonumber\\
&& \left[\!\!\begin{array}{cc}
 \mathbf{I} \!&\! \mathbf{0}
\end{array}
\!\!\right] \mathbf{A}_{c,l} \!=\! \lambda_{c,l}\bigg( \mathbf{H}_{c,l} \!-\! \left[\!\!\begin{array}{cc}
 \mathbf{0} \!&\! \mathbf{\bar{h}}_{c,l}
\end{array}
\!\!\right] \bigg).
\end{eqnarray}
Pre-multiplying $ \left[\begin{array}{cc}
\mathbf{I} & \mathbf{0}
\end{array}
\right] $ and post-multiplying $ \mathbf{H}_{c,l}^{H} $ to \eqref{eq:Another_KKTs}, it follows
\begin{equation}
\begin{split}
& \left[\!\!\begin{array}{cc}\mathbf{I} \!&\! \mathbf{0} \end{array}
\!\!\right]\mathbf{A}_{c,l}\mathbf{T}_{c,l}\mathbf{H}_{c,l}^{H}\\
& ~ - \left[\!\!\begin{array}{cc}
\mathbf{I} \!& \!\mathbf{0}
\end{array}
\!\!\right] \mathbf{H}_{c,l}^{H}\big( t_1 \mathbf{Q} \!-\! (2^{\bar{R}_{c,l}} \!-\! t_1)\mathbf{W} \big)\mathbf{H}_{c,l}\mathbf{T}_{c,l}\mathbf{H}_{c,l}^{H}  \!=\!   \mathbf{0}, \nonumber
\end{split}
\end{equation}
and thus we get
\begin{equation}
\begin{split}
 &\big( \lambda_{c,l}\mathbf{I} \!+\! t_1 \mathbf{Q} - (2^{\bar{R}_{c,l}} - t_1)\mathbf{W}  \big)\mathbf{H}_{c,l}\mathbf{T}_{c,l}\mathbf{H}_{c,l}^{H}  \\
 &= \lambda_{c,l} \left[\!\!\begin{array}{cc}
\mathbf{0} \!&\! \mathbf{\bar{h}}_{c,l}
\end{array}
\!\!\right] \mathbf{T}_{c,l}\mathbf{H}_{c,l}^{H}. \nonumber
\end{split}
\end{equation}

It is easily verified that $  \lambda_{c,l}\mathbf{I} + t_1 \mathbf{Q} - (2^{\bar{R}_{c,l}} - t_1)\mathbf{W}   \succeq \mathbf{0} $ and it is non-singular. Thus multiplying this matrix will not change the rank of the resulting matrix, and we obtain
\begin{equation}
\begin{split}
& \textrm{rank}(\mathbf{H}_{c,l}\mathbf{T}_{c,l}\mathbf{H}_{c,l}^{H}) \\
= ~&  \textrm{rank}\bigg(\!\big(\! \lambda_{c,l}\mathbf{I} \!+\!
t_1 \mathbf{Q} \!-\! (2^{\bar{R}_{c,l}} \!-\! t_1)\mathbf{W}\!\big)\mathbf{H}_{c,l}\mathbf{T}_{c,l}\mathbf{H}_{c,l}^{H}\!\bigg) \\
=~ &  \textrm{rank}\bigg(\! \lambda_{c,l} \left[\!\begin{array}{cc}
\mathbf{0} \!&\! \mathbf{\bar{h}}_{c,l}
\end{array}
\!\right] \mathbf{T}_{c,l}\mathbf{H}_{c,l}^{H} \!\bigg)\\
\leq~  &  \textrm{rank}\bigg(\! \left[\!\!\begin{array}{cc}
\mathbf{0} \!&\! \mathbf{\bar{h}}_{c,l}
\end{array}
\!\!\right] \!\bigg)  \!\leq\! 1.
\end{split}
\end{equation}
This completes the proof of Theorem 3.~~~~~~~~~~~~~~~~~~~~~~~~~~~~~~~~~~~~$ \blacksquare $
\end{document}